\documentclass[12pt]{article} 

\begin{document}


\makeatletter
\renewcommand{\@biblabel}[1]{\quad#1.}
\makeatother

\hsize=6.15in
\vsize=8.2in
\hoffset=-0.42in
\voffset=-0.3435in

\normalbaselineskip=24pt\normalbaselines


Published in: {\it Journal of Neurophysiology} {\bf 122}: 1473-1490 (2019).  

\vspace{0.25cm}

\begin{center}
{\large \bf Metabolic constraints on synaptic learning and memory}
\end{center}

\vspace{0.15cm}

\begin{center}
{Jan Karbowski$^{a,b}$}
\end{center}

\vspace{0.05cm}

\begin{center}
{\it $^{(a)}$  Institute of Applied Mathematics and Mechanics,  \\
University of Warsaw, ul. Banacha 2, 02-097 Warsaw, Poland;  \\
$^{(b)}$ Nalecz Institute of Biocybernetics and Biomedical Engineering, \\
Polish Academy of Sciences, 02-109 Warsaw, Poland }
\end{center}


\vspace{0.1cm}

\begin{abstract}

Dendritic spines, the carriers of long-term memory, occupy a small fraction
of cortical space, and yet they are the major consumers of brain metabolic
energy. What fraction of this energy goes for synaptic plasticity, correlated with
learning and memory? It is estimated here based on neurophysiological and proteomic
data for rat brain that, depending on the level of protein phosphorylation, the
energy cost of synaptic plasticity constitutes a small fraction of the energy used
for fast excitatory synaptic transmission, typically  $4.0-11.2 \%$.
Next, this study analyzes a metabolic cost of a new learning and its memory
trace in relation to the cost of prior memories, using a class of cascade models
of synaptic plasticity. It is argued that these models must contain bidirectional
cyclic motifs, related to protein phosphorylation, to be compatible with basic
thermodynamic principles.
For most investigated parameters longer memories generally require proportionally
more energy to store. The exception are the parameters controlling the speed of
molecular transitions (e.g. ATP driven phosphorylation rate), for which memory
lifetime per invested energy can increase progressively for longer memories.
Furthermore, in general, a memory trace decouples dynamically from a corresponding
synaptic metabolic rate such that the energy expended on a new learning and its
memory trace constitutes in most cases only a small fraction of the baseline
energy associated with prior memories. Taken together, these empirical and
theoretical results suggest a metabolic efficiency of synaptically stored
information.

\end{abstract}




\vspace{0.3cm}

\noindent Email:  jkarbowski@mimuw.edu.pl

\vspace{0.3cm}



\vspace{2.3cm}

\noindent {\bf Author Summary:} \\
Learning and memory involve a sequence of molecular events in dendritic spines, called
synaptic plasticity. These events are physical in nature and require energy, which has to
be supplied by ATP molecules. However, our knowledge of the energetics of these processes
is very poor. This study estimates the empirical energy cost of synaptic plasticity, and
considers theoretically a metabolic rate of learning and its memory trace in a class of
cascade models of synaptic plasticity.

\newpage

{\large \bf INTRODUCTION}

Brain is one of the most expensive organs in the body (Aiello and Wheeler 1995;
Attwell and Laughlin 2001; Clarke and Sokoloff 1994), and larger brains consume
correspondingly more energy (Karbowski 2007, 2011). Data indicate that
much of this energy is used by synapses (Harris et al 2012; Karbowski 2014).
For example, it was estimated that fast excitatory synaptic transmission requires
$1.4\cdot 10^{5}$ ATP molecules per presynaptic stimulation to pump out an influx of
Na$^{+}$ ions (Attwell and Laughlin 2001), which assuming 1 Hz for such stimulation
in the rat cortex (see below) yields a metabolic rate of $8.4\cdot 10^{6}$ ATP/min
per spine.

High metabolic requirement of synapses is also visible during mammalian brain
development when cerebral metabolic rate changes in proportion to changes in synaptic
density (Karbowski 2012). On the other hand, imaging experiments show that brain
stimulation increases cerebral metabolic rate only weakly by $\sim 10 \%$ from its resting
(baseline) value (Shulman and Rothman 1998; Shulman et al 1999), which may suggest
that housekeeping processes in the brain are more energy demanding than acquiring and
processing of a new information (Shulman et al 2004; Raichle and Mintun 2006).
Interestingly, recent data on cortical stimulation and energetics of synaptic
transmission in rodents reveal that the small increase in the cortical metabolic
rate is shared proportionally between neurons and astrocytes (Sonnay et al 2016, 2018),
implying that both neuronal and glial compartments are important for synaptic
function.

Learning and memory in the brain is strictly associated with plasticity mechanisms
in synapses (Lisman et al 2012; Kandel et al 2014; Takeuchi et al 2014; Poo et al 2016).
There exist a huge literature on modeling of synaptic plasticity and memory (for reviews
see e.g. Bhalla 2014; Chaudhuri and Fiete 2016), but virtually all of it neglects the
energetic aspect. This omission is surprising, because most of what we know about real
time brain workings is based on imaging techniques, which rely on brain metabolism
(Shulman et al 2004; Raichle and Mintun 2006). Moreover, molecular
processes in synapses are physical in nature and require a permanent energy influx to
counteract dissipation, which is related to ATP (and/or GTP) hydrolysis
(Engl and Attwell 2015; Rolfe and Brown 1997; Phillips et al 2012).
This means that there should exist some cost to learning and memory in
the brain, and an open fundamental questions is how big is it, and to what extent it
can constrain the strength and duration of a memory trace?

Synaptic plasticity mechanisms related to long term potentiation (LTP) and depression
(LTD) are induced by calcium influx to a dendritic spine through NMDA and voltage-gated
receptors, and subsequent activation of various enzymes, such as CAMKII, PSD-95, protein
kinase A and C, MAPK, etc (Lisman et al 2012; Kandel et al 2014; Bhalla and Iyengar 1999).
These enzymes serve as upstream initiators of complex molecular signaling pathways
from spine membrane to postsynaptic density (PSD), within PSD, and beyond in the form
of protein activation cascades (Sheng and Hoogenraad 2007; Zhu et al 2016). The most
common mechanism of protein activation is by phosphorylation (adding of a phosphate
group), which is powered by ATP hydrolysis (Qian 2007), and it was found that many
PSD proteins are phosphorylated during initiation of LTP and LTD
(Coba et al 2009; Li et al 2016). Thus, protein
phosphorylation cascades provide a basic biochemical mechanism of signal transduction 
in plastic synapses. The end product of the phosphorylation cascades is protein 
synthesis in PSD, actin polymerization, and AMPA receptor trafficking on spine membrane 
(Cingolani and Goda 2008; Lisman et al 2012; Bosch et al 2014). All of these processes
influence synaptic conductance/weight (Kasai et al 2003; Meyer et al 2014), and they
cause some energy drain (consumption of ATP), which magnitude is essentially unknown.
This paper provides an estimate of the metabolic cost of these reactions.

Phenomenological models of cascade synaptic plasticity mimic the richness of
biochemical pathways in dendritic spines and provide simple means to study theoretically
synaptic memory maintenance (Fusi et al 2005; Leibold and Kempter 2008; Barrett
et al 2009; Benna and Fusi 2016). Recent developments within these models
shed light on the importance of internal synaptic complexity, i.e. bidirectionality
of synaptic transitions and multiple time scales, for producing long memory 
lifetimes (Benna and Fusi 2016). The aim of this study is to consider the energy
requirement for learning and maintaining of a new information in relation to the baseline
cost of synaptic plasticity in a class of cascade models. It is argued that
these models, in order to be a minimally physically realistic, must contain bidirectional
transitions and cyclic motifs (modeled here as phosphorylation-dephosphorylation
biochemical reactions). These two features are necessary to generate nonzero and
finite synaptic metabolic rate at a steady state, corresponding to a baseline synaptic
activity, which presumably stores the memories of all previous plasticity events.
Keeping these memories is physically associated with maintaining synaptic structure and
processes, and this requires an energy influx to counteract dissipation. This implies
that synapses in the steady state or during baseline activity must operate out of thermal
equilibrium to freely exchange energy and material with their environment (neurons,
glia). This is similar to the behavior of all biological systems that have to be
in nonequilibrium state to avoid ``thermal death'' (Hill 1989; Nicolis and Prigogine 1977;
Qian 2006).


\newpage

{\large \bf RESULTS}

\vspace{0.3cm}

\noindent
{\large \bf  Estimates of energy requirements of various molecular processes
involved in synaptic plasticity.}

Synaptic plasticity of excitatory synapses, i.e. change in synaptic conductance (weight)
and size, involves a sequence of molecular events and can be broadly divided into
three categories: extra-synaptic (outside dendritic spine), intra-synaptic (inside the spine),
and modulatory. Below we estimate ATP rates associated with each of these contributions to
the plasticity based on empirical data for rat brain.

\vspace{0.2cm}
\noindent
{\it\bf Extra-synaptic cost.}

Extra-synaptic input is necessary for the initiation of synaptic plasticity. In particular,
the induction of synaptic plasticity requires the influx of Ca$^{+2}$ ions to the dendritic
spine (Miller et al 2005; Lisman et al 2012). Calcium signal serves as a trigger of various
downstream molecular pathways necessary to induce LTP and/or LTD, with the end result of
changing the spine conductance weight and size (Lisman et al 2012; Poo et al 2016). We consider
two types of the extra-synaptic costs: plasticity related glutamate recycling via astrocytes
(Sonnay et al 2016, 2018), and plasticity related ATP release from astrocytes that binds to
the spine membrane (Khan and North 2012), as the energy associated with these two processes
is directly linked to the energy cost of synaptic plasticity initiation,
i.e. calcium entering the spine.

\noindent
\underline{Plasticity related glutamate recycling.} \\
The number of glutamate molecules released by one vesicle of a presynaptic terminal
is large ($\sim 4000$ plus about 1/3 of that number released from a neighboring astrocyte;
see Attwell and Laughlin (2001) and Jourdain et al (2007)).
However, for plasticity initiation only a small fraction of this number matters, and
these are glutamates that bind directly to NMDA spine receptors to allow Ca$^{2+}$ influx.
There are roughly 10 NMDA receptors on the spine (Nimchinsky et al 2004), each
binding 1 glutamate, with a stimulation rate equal to the presynaptic firing rate times
the neurotransmitter release probability. The average firing rate in rat cortex is
about 4-5 Hz (Schoenbaum et al 1999; Fanselow and Nicolelis 1999; Attwell and Laughlin
2001; Karbowski 2009), while the release probability at these frequencies is about 0.25
(Volgushev et al 2004), which yields about 1 Hz for the rate of NMDA stimulation.
When glutamate unbinds from NMDA, it is recycled for the next use. Approximately 2.67 ATP
molecules have to be hydrolyzed for each recycled glutamate (Attwell and Laughlin 2001),
which happens mostly through astrocytes and involves several steps, such as glutamate
uptake, its metabolic processing and conversion to glutamine, glutamine transporting
to neurons, and glutamate packing into vesicles (Sonnay et al 2016, 2018). Thus in total,
the plasticity related glutamate recycling ATP rate is $10\cdot 2.67$ ATP/s or 1602 ATP/min.

\noindent
\underline{Plasticity related ATP release from astrocytes and binding to spines.} \\
Another molecule that binds to the spine receptors is ATP, which in this case plays
the role of a transmitter (Khan and North 2012). ATP released from astrocytes can
activate purinergic P2X receptors (by transducing its energy), which are known to
modulate synaptic plasticity (Pankratov et al 2009), allowing calcium to enter the
spine (see also below the section ``plasticity modulation''). It was measured that
the postsynaptic current through P2X receptors constitutes only about 10$\%$ of
the current passing through glutamate receptors, mostly AMPA (Khan and North 2012).
Since the number of AMPA on the spine is about 100 (Matsuzaki et al 2001), this suggests
that the number of P2X is about 10 (P2X and AMPA channels have comparable conductances;
Khakh and North 2012). Each P2X binds 3 ATP molecules for its activation (Bean et al 1990).
The rate of ATP release from astrocytes is not known (it is likely activity dependent),
but we can assume that the timing between two consecutive releases should be at least as
long as the ATP binding time constant, which for most of P2X receptors is long and more than
20 sec. Assuming 30 sec, this gives us 30 released ATP molecules related to plasticity per
30 sec, or the consumption rate of 60 ATP/min.

\vspace{0.2cm}
\noindent
{\it\bf Intra-synaptic cost.}

Below we estimate the energy requirements for the basic molecular processes
inside a dendritic spine, which are related to plasticity induction and maintenance.
These include: Ca$^{2+}$ intra-spine trafficking, protein phosphorylation, protein
synthesis (turnover), actin treadmilling, and receptor (AMPA and NMDA) trafficking.

\noindent
\underline{Ca$^{2+}$ removal.} \\
Following a presynaptic action potential or postsynaptic backpropagation calcium
flows into the spine (either through NMDA channels or voltage sensitive Ca$^{2+}$
channels) and trigger a cascade of molecular processes, initiating the synaptic
plasticity. For a spine volume of 0.1 $\mu$m$^{3}$ (Honkura et al 2008), it was found
that about 2000 Ca$^{2+}$ ions enter, of which about 95$\%$ bind to endogenous buffers
(Sabatini et al 2002). The remaining 100 Ca$^{2+}$ are free and can activate CaMKII
(and possibly other) proteins, but they have to be pumped out after the activation to
recover baseline physiological conditions of resting calcium concentration in the spine.
The rate of the extrusion is given by presynaptic stimulation (1 Hz), and is conducted by
Ca$^{2+}$-ATP pumps or Na$^{+}$-Ca$^{2+}$ exchanger. Since both of these processes hydrolyze
1 ATP molecule per 1 removed Ca$^{2+}$ ion, we find the ATP hydrolysis rate of calcium
extrusion from the spine 100 ATP/s or 6000 ATP/min.


\noindent
\underline{Protein phosphorylation.} \\
Protein phosphorylation is the main activation mechanism of downstream proteins,
actin, AMPA receptors, and other spine molecules, and thus it is of prime importance
(Zhu et al 2016). One cycle of protein phosphorylation requires the hydrolysis of
1 ATP molecule (Hill 1989; Qian 2007). First, we estimate the ATP consumption rate
for resting non-LTP related phosphorylation, i.e. for unstimulated spine with resting
Ca$^{2+}$ concentration, and then for stimulated spine undergoing LTP.

In general, for the resting spine the levels of protein phosphorylation rates seem to
be uniformly distributed and vary by two orders of magnitude, from 0.001 min$^{-1}$
to 0.34 min$^{-1}$ (Molden et al 2014), which yields an average value of 0.15 min$^{-1}$.
There are about $10^{4}$ proteins (including their copies) in the spine PSD
(Sheng and Kim 2011), with an average of 4-6 phosphorylation sites per protein
(Collins et al 2005; Trinidad et al 2012). This gives the resting rate of ATP-driven
protein phosphorylation as $(6-9)\cdot 10^{3}$ ATP/min, with an average of 7500 ATP/min.
Because hydrolysis of one ATP requires 20 kT, where k is the Boltzmann constant and T
is the absolute temperature (Phillips et al 2012), we obtain equivalently the resting
energy rate of protein phosphorylation as $15\cdot 10^{4}$ kT/min in a single
spine. This ATP rate is however unrelated to learning and memory, since the resting
conditions (unstimulated spine) do not drive plasticity events, i.e., changes in AMPA
receptor number.

More relevant rates for LTP-related ATP consumption can be obtained by noting
that during the plasticity induction, stimulated by Ca$^{2+}$ influx to the spine,
the phosphorylation rates of about $10\%$ PSD proteins are strongly enhanced as the proteins
interact more frequently (about 130 proteins out of roughly 1500;
Coba et al 2009; Li et al 2016; Bayes et al 2012). The important point is that
this fraction is dependent on the frequency of Ca$^{2+}$ stimulation
(De Koninck and Schulman 1998; Gaertner et al 2004). For example, the rate of CaMKII
autophosphorylation jumps $3-4$ orders of magnitude, from the resting of 0.03 min$^{-1}$
(Colbran 1993; Miller et al 2005) to about $60-600$ min$^{-1}$ 
(Bradshaw et al 2002; Miller et al 2005; Michalski 2013), but this amplified phosphorylation
takes place only for a few percent of CaMKII at about 1 Hz of Ca$^{2+}$ influx
(De Koninck and Schulman 1998; Gaertner et al 2004). We can expect that during continuing
calcium stimulation, as for regular {\it in vivo} cortical conditions, some balance between
proteins highly phosphorylated and those at resting phosphorylation is achieved. In such
a stationary state, both of these protein groups contribute to the ATP consumption rate,
which can be written as:

$\dot{ATP}_{phos}= N[(1-x)r_{r}M_{s} + xr_{a}M_{s}^{*}]$,

where $N (= 10^{4})$ is the total number of proteins (including their copies) in PSD,
$x (=0.1)$ is the fraction of proteins with amplified phosphorylation, $M_{s} (=5)$ is
the average number of phosphorylation sites per protein, $M_{s}^{*}$ is the average
number of such sites per protein that become highly phosphorylated upon stimulation
($M_{s}^{*}$ can be any integer between 1 and 5). $M_{s}^{*}$ depends on the frequency
of Ca$^{2+}$ stimulation, and it likely assumes its lowest values $\approx 1-2$
(De Koninck and Schulman 1998; Gaertner et al 2004).
Finally, $r_{a}$ and $r_{r}$ denote the average rates for active and resting
phosphorylation (population averaged resting $r_{r}=0.15$ min$^{-1}$). There are virtually
no data on $r_{a}$ in PSD proteins other than CaMKII. For this reason, in our estimate we
take for $r_{a}$ a medium value of the numbers reported for CaMKII, i.e., $r_{a}\approx 300$
min$^{-1}$. Using the above parameters, we obtain the steady-state ATP consumption rate
during LTP phase as $\dot{ATP}_{phos}= 3.1\cdot 10^{5}$ ATP/min for $M_{s}^{*}= 1$, and
$\dot{ATP}_{phos}= 9.1\cdot 10^{5}$ ATP/min for $M_{s}^{*}= 3$. These values correspond
respectively to 20$\%$ and 60$\%$ of the protein sites with enhanced phosphorylation.

As a curiosity, let us estimate the above $\dot{ATP}_{phos}$ for different values of the
fraction of highly activated proteins $x$. First, note that for $x= 0$ (no proteins
with amplified phosphorylation), the $\dot{ATP}_{phos}$ is exactly equal to the resting
non-LTP phosphorylation cost calculated above. Next, since $\dot{ATP}_{phos}$ increases
with $x$, we want to find out for what value of $x$ the ATP cost of protein phosphorylation
is equal to the cost of fast synaptic transmission ($8.4\cdot 10^{6}$ ATP/min)?
For $M_{s}^{*}=1$ this never happens, and the maximal value of $\dot{ATP}_{phos}$ in
this case is $3\cdot 10^{6}$ ATP/min, which is 36$\%$ of the fast synaptic transmission
cost. For $M_{s}^{*}= 3$ this happens when $x=0.94$, i.e., almost all PSD proteins
would have to be activated by phosphorylation. For the maximal value of
$M_{s}^{*}= 5$, this situation occurs for $x=0.56$, i.e. half of the proteins must
be highly activated. Finally, the maximal possible value of $\dot{ATP}_{phos}$ is 
$15\cdot 10^{6}$ ATP/min (all PSD proteins are maximally phosphorylated), which is
nearly twice the cost of fast synaptic transmission. The latter hypothetical calculation
is useful, because it sets the upper bound on the possible error in estimating the
overall cost of synaptic plasticity.

It is interesting to estimate what fraction of the resting global phosphorylation
cost is taken by two the most abundant proteins: CaMKII ($\alpha$ and $\beta$ subunits)
and PSD-95. There are 5600 copies of CaMKII$\alpha$ and CaMKII$\beta$ in a spine
(Sheng and Kim 2011), and both subunits have similar number of 10 phosphorylation sites
(Trinidad et al 2006). The number of copies of the PSD-95 protein in a spine is 300
(Sheng and Kim 2011), and it has between 8 and 12 phosphorylation sites (Trinidad et al
2006; Zhang et al 2011). Assuming that both proteins have the same resting phosphorylation
rates, i.e., 0.03 min$^{-1}$, corresponding to CaMKII autophosphorylation, we get
the energy rate 1700 ATP/min for CaMKII, and $60-100$ ATP/min for PSD-95.
If we combine these two numbers, we get that these two proteins consume $20-30 \%$
of the global resting phosphorylation energy rate of the whole postsynaptic density PSD.
This indicates that these two plentiful proteins, comprising about 60$\%$ of the PSD
content, consume a substantial part of the whole energy devoted to protein phosphorylation
during resting non-LTP spine conditions.
 
\noindent
\underline{Protein synthesis/turnover.} \\
The cost of protein turnover is estimated as follows. The total molecular mass of
a typical PSD has been calculated as $1.1\cdot 10^{9}$ Da (Chen et al 2005), and this
corresponds to a total number of $10^{7}$ amino acids, which are bound by the same
number of peptide bonds that require 4 ATP molecules/bond to form (Engl and Attwell
2015). The average half-lifetime of PSD proteins is 3.67 days (Cohen et al 2013),
which means that after that time a half of all peptide bonds are broken. This means
that the ATP consumption rate for protein synthesis as $3.7\cdot 10^{3}$ ATP/min
(or equivalently $7.4\cdot 10^{4}$ kT/min) for a single spine.

Similarly, we also estimate the fraction of global protein synthesis cost taken by
CaMKII and PSD-95. Two subunits CaMKII$\alpha$ and CaMKII$\beta$ have similar molecular
masses with an average length of 528 amino acids, while PSD-95 has 779 amino acids
(Yoshimura et al 2004). The half-times for CAMKII decay is 3.4 days (average of 3.0
for CaMKII$\alpha$ and 3.8 for CaMKII$\alpha$) and for PSD-95 decay is 3.67 days
(Cohen et al 2013). Given that there are 5600 copies of CaMKII and 300 copies of
PSD-95 (Sheng and Kim 2011), this yields a synthesis rate 0.57 copies/min of CaMKII,
and 0.03 copies/min of PSD-95. This requires $4\cdot 528\cdot 0.57= 1200$ ATP/min for
CaMKII and $4\cdot 779\cdot 0.03= 93$ ATP/min for PSD-95. From this, it follows that
these two the most frequent proteins consume for their synthesis about 1/3 of the total
energy used for protein synthesis in the whole PSD.

\noindent
\underline{Actin treadmilling.} \\
Actin treadmilling energetics in spines can be estimated as follows. Each cycle of
actin treadmilling involves 1 ATP. Actin concentration in the mammalian brain is
100 $\mu$M (Devineni et al 1999), which yields 6000 actin molecules per spine
(average spine volume assumed: 0.1 $\mu$m$^{3}$ (Honkura et al 2008)). About 90 $\%$
of actin in spines degrades fast with a characteristic lifetime $\sim 40$ sec,
and the remaining 10 $\%$ with much longer time $\sim 17$ min
(Honkura et al 2008; Star et al 2002), which can be neglected. This means that
the metabolic cost of actin turnover is 8100 ATP/min (or $1.6\cdot 10^{5}$ kT/min).
   
\noindent
\underline{AMPA and NMDA trafficking.} \\
The energy cost of AMPA and NMDA receptor trafficking within a spine is composed
of receptor insertion/removal to/from the spine membrane (Huganir and Nicoll 2013)
and receptor movement along the membrane (Choquet and Triller 2013).
First, we consider the receptor insertion (exocytosis) and removal
(endocytosis) contributions. Both of these processes can be envisioned as molecular
crossing of energy barriers, because both of them lead to deformations in the
membrane structure. The interesting point is that the rates of endo- and exocytosis
of AMPA receptors are very similar at steady state and about 0.1 min$^{-1}$ (Ehlers 2000,
Lin et al 2000), which is much faster than the turnover (degradation) rates for
AMPA (half-life about 2 days; Cohen et al 2013). The approximate equality of the
insertion and removal rates indicates that the energy barriers for these two opposing
processes are similar (invoking the Arrhenius law). A typical energy barrier for
AMPA insertion is $4-30$ kT, with an average of 17 kT, which is the energy of protein
insertion into a lipid membrane by a mechanism of membrane fusion (Grafmuller et al 2009;
Gumbart et al 2011; Francois-Martin et al 2017). Since an average spine contains about
100 AMPA (Matsuzaki et al 2001), we obtain the energy rate for AMPA endocytosis/exocytosis
as $2\cdot 100(17 kT)\cdot 0.1$ min$^{-1}$ (the prefactor 2 comes from including both
endo- and exocytosis transitions), which yields 340 kT/min, or equivalently 17 ATP/min.
Inclusion of NMDA receptors affects this figure by only $10\%$, since the number of NMDA
receptors on spine membrane is much lower, about 10 (Nimchinsky et al 2004).
Thus, the total cost of AMPA and NMDA insertion/internalization is 18.7 ATP/min.

Energy requirement of receptor movement is estimated
assuming that most of it is done along spine membrane, which might be an underestimate
given that some trafficking might also take place internally along actin filaments
(Choquet and Triller 2013). In general, the motion of molecules along some substrate
is powered by ATP hydrolysis (Bustamante et al 2005), with a generation of a propulsion
force $F$ of the order of 6 pN (Visscher et al 1999). Additionally, it was found that
AMPA motion along spine membrane was alternating between the periods of diffusion and
quietness (Borgdorff and Choquet 2002) with comparable durations, and a diffusion
coefficient $D= 0.02-0.05$  $\mu$m$^{2}$/s, or its average value $D= 0.035$  $\mu$m$^{2}$/s.
The power $P$ dissipated by one receptor is proportional to the product of the force $F$
and an average velocity $v$, which is $v\sim L/\tau$, where $L$ is the average spine
length and $\tau$ is the typical time to travel distance $L$. Since for diffusive processes
$\tau\sim L^{2}/D $ (Phillips et al 2012), we get that velocity $v\sim D/L$.
Thus, we can write that the power dissipated by a single receptor trafficking along spine
membrane is $P\approx 0.5 FD/L$, where the prefactor 0.5 comes from the assumption that the
time intervals of diffusion and quietness are roughly similar (Borgdorff and Choquet 2002),
which reduces the effective velocity by half. For a typical spine length of $L= 1 \mu m$,
we obtain $P\approx 1.05\cdot 10^{-19}$ J/s, which using the fact that kT is $4.23\cdot 10^{-21}$
J at 36 $^{o}C$, yields $P\approx 24.8$ kT/s, or $74.4$ ATP/min.
Given that there are 100 AMPA and 10 NMDA receptors in a typical spine
(Matsuzaki et al 2001; Nimchinsky et al 2004), we find the energy rate related
to receptor movement as 8184 ATP/min (or $1.6\cdot 10^{5}$ kT/min). 
This figure is 437 times larger than the one for receptor insertion
and internalization, and hence it is much more important.

\vspace{0.2cm}
\noindent
{\it\bf Plasticity modulation cost.}

There are many routes for modulation of synaptic strength. One of the
better studied pathways is through purinergic P2X receptors (Pankratov et al 2009;
Khakh and North 2012). Activated by ATP molecules P2X receptors, which are voltage
gated nonselective channels, enable Ca$^{2+}$ ions to flow inside the spine. This
calcium influx modulates various internal molecular pathways, with the end result
of affecting the number of AMPA receptors on the spine membrane, likely by modulating
their endo- and exocytosis rates (Gordon et al 2005; Pougnet et al 2014). Both LTP and
LTD had been observed, depending on which pathway is influenced, but the relative changes
in the synaptic strength did not exceed $30-40 \%$ (Gordon et al 2005; Pougnet et al 2014). 
This means that the rates of AMPA insertion and removal should be also affected by
this percentage, which suggests that the overall energy rate for AMPA trafficking
can increase by about $35 \%$ due to the actions of P2X receptors. This translates
to an additional 2871 ATP/min. Because of other possible pathways modulating
synaptic plasticity, e.g., by adenosinergic receptors, this figure should be considered
as a minimal amount of ATP rate due to plasticity modulation.

\vspace{0.2cm}
\noindent
{\it\bf Summary of the molecular costs.}

From these consideration it follows that the intra-synaptic processes are the most energy
demanding for the plasticity. Within them, the cost of protein phosphorylation dominates
over the rest by a factor $10-30$. The cost of the remaining intra-synaptic processes
(Ca$^{2+}$ removal, actin treadmilling, receptor trafficking) is very similar, with the
cost of protein synthesis about two times lower (Table 1).
Together, all the synaptic plasticity processes considered here use $(3.4-9.4)\cdot 10^{5}$
ATP/min and account for only about $4-11 \%$ of the metabolic cost related to the fast
excitatory synaptic transmission for rat brain (Table 2). Overall, the small values of
these percentages indicate that synaptic plasticity is metabolically inexpensive.

\newpage

\noindent
{\large \bf  Modeling the metabolic cost of learning and memory.}

\vspace{0.3cm}

The next major goal is to study theoretically the energy rate associated with a particular
form of learning and memory, and relate it to the lifetime of a new memory.

\vspace{0.2cm}

{\it \bf Motivation for the approach.} 

All the processes considered above and associated with synaptic plasticity constitute
a physical backbone of synaptic learning and memory. We want to quantify the metabolic
cost of a new learning and its memory trace, using a modeling approach. To do this, one
should in principle include all the above processes in a model. However, combining all
of them in a single model is extremely difficult and perhaps even infeasible, because it
is not clear how all these plasticity related mechanisms are mutually coupled and with
what strength. Instead, we focus
in the modeling part on the initial phase of synaptic plasticity, called induction
of plasticity (or early LTP), during which synapses start to grow and increase their
strength, which is associated with learning of a new input and its subsequent decay.
The induction phase is accompanied by enhanced phosphorylation of PSD proteins,
including AMPA receptors (while trafficking into the spine membrane), which is evident
in Table 1. This indicates that protein phosphorylation is the most energetically
demanding and dominating process during the plasticity initiation. For this reason,
and because protein phosphorylation is the most commonly studied cellular microcircuit
(Krebs 1981; Hill 1989; Qian 2007), it is the main component of the modeling part.
We neglect in the model the late phase of plasticity (plasticity maintenance or late LTP)
during which most of the protein synthesis takes place, because it uses much less energy
(Table 1). We also neglect in the model the other processes from Table 1, which means
that the actual metabolic cost of learning and memory could be slightly greater than
theoretically calculated below.

The model used for quantifying the energetics of a memory trace and its duration is
based on the so-called cascade models of synaptic plasticity (Fusi et al 2005; Leibold
and Kempter 2008; Barrett et al 2009; Benna and Fusi 2016).
In short, these models assume that synaptic molecular machinery can be described
in terms of discrete probabilistic states, with transitions between the states
depicting molecular transformations. However, the classic cascade models can not
describe properly protein phosphorylation due to their simplistic topology, i.e.,
too simple pattern of molecular transitions. Below, the cascade model of plasticity
is extended to include protein interactions via phosphorylation-dephosphorylation
mechanism, by adding transition motifs with closed loops (Qian 2007).

The expenditure of energy in the model with phosphorylation-dephosphorylation cycles
is quantified using the concept of entropy production rate EPR (Hill 1989; Nicolis
and Prigogine 1977). In our case, EPR measures the rate of dissipated energy (metabolic
rate) in the spine due to transitions between different molecular states (Rolfe and
Brown 1997; Qian 2006). When a synapse is in thermodynamic equilibrium with its
environment, no energy influx enters the synapse, all internal molecular reactions
are balanced (forward and backward reaction rates are equal), and the synapse does
not dissipate any energy, implying a vanishing metabolic rate and $\mbox{EPR}= 0$.
Such a condition corresponds to a ``thermal death'' when no biological function can
be performed, and thus it cannot represent a baseline or resting synaptic state,
during which the synapse keeps track of its prior plasticity events, i.e. keeps
their memory. The functional baseline synaptic state is certainly a driven state
that requires energy (and material) exchange with the surrounding, and this implies
that the synapse must be an open system in thermodynamic nonequilibrium even during
its steady state baseline activity.
This steady state or baseline must obviously dissipate energy, as the incoming
energy flux breaks the balance in the forward and backward rates of the internal
molecular reactions (in our case phosphorylation rates), which leads to nonzero
EPR and dissipation, as well as to finite synaptic metabolic rate. We will call
this baseline synaptic state, the nonequilibrium steady state (NESS), in analogy
to the systems considered in nonequilibrium thermodynamics (Lebowitz and Spohn 1999;
Mehta and Schwab 2012). Our theoretical NESS state corresponds to the empirical
active steady state phosphorylation phase analyzed above as the intra-synaptic
cost of plasticity.

When a synapse is stimulated (by Ca$^{2+}$ influx) and the plasticity event initiated,
the molecular reactions/transitions are amplified, which leads to perturbations in the
distribution of synaptic states and the emergence of the time dependent memory trace
associated with this perturbation. The synaptic perturbation also causes an increase in
the rate of energy dissipation (EPR), which declines after some time to its baseline
level when the stimulation is turned off. The key relationship that we want to explore,
is the one between the total energy expanded on memory trace and the memory duration.

\vspace{0.2cm}

{\it \bf Thermodynamically realistic model of cascade synaptic plasticity must 
contain cyclic reactions.}

We consider synaptic plasticity as transitions between multiple discrete synaptic
states (Montgomery and Madison 2004) (Fig. 1). These states represent internal
degrees of freedom of the molecular processes in a dendritic spine. It is assumed
that many states correspond to one synaptic weight, either weak or strong, denoted as
up and down (with the symbols $+$ and $-$ in Fig. 1C), and this reflects
neurophysiological data on a single synapse level (Petersen et al 1998; O'Connor et al 2005).
The vertical transitions between the states in Fig. 1C correspond to conventional
plasticity, while the horizontal transitions are related to the so-called
metaplasticity (Abraham and Bear 1996). The latter transitions do not lead to
changes in the synaptic weight. It turns out that not every configuration of these
transitions produces nonequilibrium steady state (NESS), required for nonzero
metabolic rate. For instance, a ``ladder'' structure of synaptic molecular reactions
(Fig. 1A) yields a zero EPR (and metabolic rate) at steady state
(see Eq. 15 and below in the Methods), which is not realistic. The basic requirement
for NESS and nonzero EPR is the presence of cyclic and bidirectional molecular reactions,
which generate nonzero probability flux that mimics the exchange of energy with an
environment (Fig. 1B; Qian 2006).

This requirement can be verified for a simple and  minimally realistic model of only 3 states
linked by a closed loop of reactions, which represents a phosphorylation-dephosphorylation
cycle for one protein interaction (Fig. 1B). Such a loop is the basic metabolic motif
for modeling synaptic plasticity with multiple states, i.e., with many phosphorylation
events (Fig. 1C). In this simple 3 state case, $p_{-}$ denotes the probability of a protein
in a ground state and $p_{+}$ is the probability of the activated protein (Fig. 1B).
The protein can go from the state $p_{-}$ to $p_{+}$ either directly, but it is rare and
occurs with a small transition rate $\epsilon_{1}\alpha$ (where $\epsilon_{1} \ll 1$), or
it can go to the state $p_{+}$ indirectly, in two steps, through an intermediate state
$q_{0}$ (by binding an enzymatic substrate). The interesting point is that this second
step, from $q_{0}$ to $p_{+}$ can be very fast, with a large transition rate $a$, depending
on the level of protein phosphorylation that is powered by ATP hydrolysis. This simple
3 state protein is not in thermal equilibrium with its environment, because it continually
transfer between the 3 states ($p_{-}$, $q_{0}$, $p_{+}$), driven by energy provided by
ATP. Therefore, this configuration dissipates energy even in the baseline, which can be
found explicitly as entropy production rate $\mbox{EPR}_{0}$ at NESS (Fig. 1B)

\begin{eqnarray}
\mbox{EPR}_{0}= kT \frac{\alpha\beta}{Z}\left(\epsilon_{2}a-\epsilon_{1}\epsilon_{3}b\right)
\ln\left(\frac{\epsilon_{2}a}{\epsilon_{1}\epsilon_{3}b}\right),
\end{eqnarray}\\
where $Z$ is a function of different transition rates $a, b, \alpha, \beta$,
with $\epsilon_{i}$ unitless small coefficients ($\epsilon_{i} \ll 1$) (see Eq. 11 in Methods).
The value of $a$ is controlled directly by protein phosphorylation rate driven by ATP hydrolysis,
and $b$ is the dephosphorylation rate.
The term  $\epsilon_{2}a-\epsilon_{1}\epsilon_{3}b$ represents the bias or deviation
from thermal equilibrium, and corresponds to the probability flux (Eq. 9) that circulates
between the three states. Eq. (1) implies that if the transitions do not form a loop, i.e.
when $a= b= 0$, then EPR$_{0}$ = 0 (since $x\ln(x) \mapsto 0$ as $x \mapsto 0$).
Moreover, to get a finite EPR$_{0}$ one needs bidirectional transitions.
For unidirectional transitions, we obtain EPR$_{0} \mapsto \infty$, as can be seen e.g. by
setting $a > 0$ and $b= 0$. This obviously is unrealistic.

The steady state EPR$_{0}$ for the 3 state system (Fig. 1B) is also 0, when the condition of
the so-called detailed balance is met, i.e., when the bias
$\epsilon_{2}a - \epsilon_{1}\epsilon_{3}b= 0$,
which corresponds to thermodynamic equilibrium. The detailed balance can be broken
(leading to nonequilibrium) if, e.g., $\epsilon_{3}$ is much smaller than the rest of
the parameters in Eq. (1). This situation happens in a living cell, where there is a big
asymmetry between phosphorylation and dephosphorylation, i.e. $a \gg \epsilon_{3}b$
(Qian 2006). Generally, the higher that asymmetry the larger EPR$_{0}$.

In the remaining of this study we investigate the metabolic constraints on learning
an input and on its subsequent memory trace, in the model with multiple states (Fig. 1C).

\vspace{0.35cm}

\noindent
{\it \bf Metabolic cost of baseline synaptic plasticity is insensitive on
  the number of synaptic states but it is affected by the transition rates
  between the states.} 

When synapses are not stimulated, their dynamics converge into baseline activity, which
is the nonequilibrium steady state NESS with distributed occupancies of different states
and multiple transition loops (Fig. 1C; Methods). All the transitions between these states
are appropriately rescaled by a prefactor $e^{-zk}$ to progressively slow down the downstream
dynamics of the loops with the higher index $k$, where $z$ is the slowing down factor.
This prefactor is introduced to provide multiple time scales in the dynamics of intrasynaptic
molecular interactions. In the NESS state in Fig. 1C, energy is constantly dissipated due to
many phosphorylation-dephosphorylation loops, and this leads to nonzero basal synaptic metabolic
rate. Entropy production rate EPR$_{0}$ associated with this state can be viewed as a metabolic
cost of maintaining all the prior plasticity events, i.e., the memory of the past, and it is
always greater than zero, and could be substantial even for one molecular pathway $\sim 0.5$
kT/min (Fig. 2). Generally, the baseline EPR$_{0}$ is essentially independent
of the number of synaptic states $n$, but it depends on the magnitude of the transition
rates between synaptic states (Fig. 2). Specifically, increasing the rate of synaptic
slowing down $z$ makes EPR$_{0}$ monotonically smaller with a saturation for large $z$.
On the other hand, EPR$_{0}$ as a function of ATP driven transitions with an amplitude
$a_{0}$ exhibits more complicated shapes: either it increases up to a saturation with
increasing $a_{0}$ for $z=0$, or it has maxima for $z > 0$ (Fig. 2). These dependencies
indicate that although the baseline NESS state requires some energy for its maintenance,
its cost can be reduced by globally slowing the transitions between synapses (increasing $z$)
and, simultaneously, by keeping the amplitude of ATP driven transitions ($a_{0}$)
either very small or very large.

\vspace{0.35cm}

\noindent
{\it \bf Stronger and longer synaptic stimulations lead to longer memory traces
 with higher energy expenditures, but with low relative costs.} 

Next, we study the energetics and the memory trace related to a single transient plasticity
event. The plasticity event corresponds to synaptic stimulation by affecting all ATP-driven
phosphorylation rates $a_{k}^{\pm}$ associated with the transitions $q_{k}^{\pm} \mapsto p_{k+1}^{\pm}$,
which is a microscopic representation of some macroscopic learning (Fig. 1C). The stimulation
is induced by a brief jump in the ATP-driven transition rates $a^{\pm}_{k}$, by a relative fraction
$A_{\pm}(t)= A_{\pm}\exp(-t/\tau)$, which relaxes exponentially to zero with a time constant $\tau$
that can be also thought as a learning time. This perturbation leads to a redistribution of
the occupancy probabilities of the synaptic states, and the recovery to the baseline probabilities
takes some time, called memory lifetime $T_{m}$, which in general is much longer than the
stimulation time $\tau$.

The memory trace associated with the single stimulation is defined as the deviation of the
average synaptic weight from its baseline value, relative to a noise in the weights. This
essentially corresponds to a signal-to-noise ratio SNR, which is given by Eqs. (12-14) in
the Methods. Temporal dependences of memory trace SNR and synaptic energy rate EPR following
a single stimulation are presented in Fig. 3. Upon stimulation from the NESS state, SNR
initially builds up to some maximal value and then it slowly decays with a characteristic
long tail, given by SNR$\sim t^{-\delta}$, where $\delta\approx 4/3$ (Fig. 3). Memory trace
is detectable if SNR is above a certain threshold, which is taken here as 1 (the results are
insensitive on the precise value of this threshold), and the corresponding memory lifetime $T_{m}$
is defined as the time interval from the stimulation up to the moment when SNR drops below the
threshold. Energy rate EPR associated with this stimulation has a qualitatively different
time course than SNR, as it follows closely the temporal dependence of short-term
stimulation $A_{\pm}(t)$, without exhibiting a long tail (Fig. 3). This means that EPR
starts from a high level and quickly decays to its resting value EPR$_{0}$ when the
stimulation ends (Fig. 3). Thus, most of the time when memory trace is still detectable,
the energy rate is essentially at its resting value, which implies that the two variables
decouple for longer times (Fig. 3). A direct consequence of this important observation is
that the total energy $E$ expanded on a new memory trace, defined as an area
under EPR, differs in most cases only by a small percentage from a baseline energy
$E_{0}$ ($E_{0}= \mbox{EPR}_{0}T_{m}$) required for supporting the baseline synaptic state
related to all prior memories during time $T_{m}$ (Fig. 4; see also below).
This effect is much more pronounced for larger slowing down $z$, when the speed of
downstream molecular reactions is severely reduced (Fig. 4). The relatively low energy
cost is a sign of metabolic efficiency of a new memory storing in the cascade model of
synaptic plasticity.

\vspace{0.35cm}

\noindent
{\it \bf When longer memories require proportionally larger energy expenditure?} 

How the lifetime $T_{m}$ of the memory trace and its energy cost $E$ depend on the
amplitude $A_{+}$ and duration $\tau$ of synaptic stimulation?
Memory lifetime $T_{m}$ generally increases with increasing $A_{+}$ and $\tau$, but both
dependences are roughly logarithmic (Fig. 5). The corresponding energy expenditure $E$
on keeping the new memory trace SNR above the threshold also grows with $A_{+}$ and $\tau$
in a similar manner as $T_{m}$ (Fig. 5). Interestingly, the energy cost $E$ of a new memory
increases in proportion to its lifetime $T_{m}$, as is evident by an approximate constancy
of the ratio $T_{m}/E$ over variability in the stimulus amplitude and its duration (Fig. 5).
Moreover, the ratio $T_{m}/E$ is larger for larger $z$, which implies that the gain in
memory duration per invested energy is bigger if the rates of downstream molecular processes
are reduced.

How general is the finding that energy cost of a memory trace increases proportionally
with the memory lifetime? What happens if we keep the level of synaptic stimulation constant,
and instead, change the intrinsic parameters characterizing synaptic plasticity?
In Figs. (6-7) we show the dependence of $T_{m}$ and an associated energy cost $E$
on two internal synaptic parameters: basic number of synaptic states $n$, and the fraction
of potentiated synapses $f$.

Increasing synaptic states $n$ leads to an increase in $T_{m}$, but only for small $n$
(Fig. 6). For larger $n$, the memory lifetime $T_{m}$ saturates. The related energy cost
$E$ behaves similarly, such that $E$ is proportional to $T_{m}$, which again follows from
the observation that the ratio $T_{m}/E$ does not change much over the whole range of $n$.
A qualitatively similar pattern, with approximate constancy of $T_{m}/E$, is observed when
the fraction of potentiated synapses $f$ is changed (Fig. 7).

Taken together, in all these three cases a longer memory trace requires a proportionally
more energy to sustain.

\vspace{0.35cm}

\noindent
{\it \bf When longer memories do not require more energy?} 

A different picture emerges, regarding the relationship between $E$ and $T_{m}$, when two
other internal parameters are changed that are related to the speed of molecular transitions.
One is the molecular slowing-down factor $z$, and another is the global amplitude of the
phosphorylation rate $a_{0}$.

Growing the parameter $z$ leads to bimodal shapes of memory trace duration $T_{m}$ and
its energy cost $E$ with the appearance of maxima (Fig. 8). In this case, however, the ratio
$T_{m}/E$ increases significantly with growing $z$, which indicates a substantial gain in
memory duration per expanded energy (Fig. 8). This means that memory lifetime grows
faster than its energy cost, i.e., longer memories are relatively cheaper.
Interestingly, the ratio $T_{m}/E$ is insensitive to the fraction of potentiated
synapses $f$, as all dependencies collapse on a single line (Fig. 8).

A slightly more complex scenario appears when the ATP-driven phosphorylation amplitude
$a_{0}$ is varied (Fig. 9). For $z= 0$, the memory lifetime $T_{m}$ and energy $E$ both
increase monotonically with $a_{0}$ such that the ratio $T_{m}/E$ initially decreases and
then stabilizes at some level. For a more interesting case $z > 0$, the memory duration
$T_{m}$ and its cost $E$ are not easily correlated: $T_{m}$ initially grows and then saturates
for larger $a_{0}$, while the associated energy cost $E$ always exhibits a maximum at some
$a_{0}$. More importantly, for $z > 0$, the ratio $T_{m}/E$ displays a minimum, which is very
close to the point where energy $E$ has a maximum (Fig. 9). This means that there are two
different regimes: for small $a_{0}$ a relative cost of increasing memory lifetime strongly
grows (sharp decrease in the ratio $T_{m}/E$), whereas for large $a_{0}$ the opposite happens
and the relative cost of memory duration decreases ($T_{m}/E$ increases). Thus, the latter
regime is much more energy efficient, which is also visible in the high values of both
$T_{m}$ and $T_{m}/E$ for large $a_{0}$ (Fig. 9).

Taken together, these two results suggest that storing longer memories is not always
associated with a higher metabolic burden. In fact, there can be regimes in the internal
synaptic parameters, here $z$ and $a_{0}$, for which longer memories can be
relatively cheap.

\vspace{0.35cm}

\noindent
{\it \bf  Metabolic cost of a new memory relative to the cost of prior memories.} 

How expensive is to invoke a new plasticity event and keep its memory in the molecular
interactions, relative to the cost of prior plasticity events ``encoded'' in the baseline
spine metabolic rate? Figure 4 as well as lower panels of Figs. (6-9) provide an answer
to this question. The relative cost of a new memory trace with respect to the baseline
energy cost $E_{0}$ during $T_{m}$, i.e. the ratio $(E-E_{0})/E_{0}$, is almost always
smaller or much smaller than 1. Thus, the cost of keeping the new memory trace detectable
is generally marginal to the cost of storing memories of all previous events.

\newpage

{\large \bf DISCUSSION}

\vspace{0.35cm}

\noindent
{\large \bf Empirical metabolic efficiency of long-term synaptic plasticity.}

Dendritic spines of excitatory synapses occupy only about 10$\%$ of neocortical volume
of adult mammalian brain (Karbowski 2015), but their short-term signaling associated
with fast synaptic transmission can cause a high metabolic burden to the whole cortex.
Its estimated cost in the rat cortex is $8.4\cdot 10^{6}$ ATP/min per spine (Attwell and
Laughlin 2001), and the interesting question is how does it relate to the synaptic
plasticity cost?

Proteins underlying molecular level of synaptic learning and memory constantly interact,
which is associated with synaptic plasticity, and these processes require energy influx
and thus some metabolic cost (Table 1). What is this overall cost, and to what extent it can
constrain the memory trace? The empirical estimates conducted here suggest that the energy
related to synaptic plasticity in rat cortex comprises only $4-11 \%$ of the energy used by
excitatory synapses for fast synaptic transmission given above (Table 2). Considering
that the latter energy (including all spines) constitutes around $30-80 \%$ of the total cortical
metabolic rate (Attwell and Laughlin 2001; Harris et al 2012; Karbowski 2009 and 2012),
we get that metabolic cost of learning and memory is only about $1-9 \%$ of the total
metabolic rate. This is the empirical evidence that the processes of learning and information
storing in synapses are energetically rather cheap.

What about the energetic constraints on memory trace? The cascade model considered here
suggests that, although a longer memory trace requires proportionally larger amounts
of energy (in most cases; Figs. 5-7), these amounts are relatively small ($ < 10\%$)
as compared to the baseline synaptic plasticity cost (Figs. 6-9).

Thus, it seems that synaptic plasticity is non-demanding metabolically in the scale of
the whole brain, but nevertheless it can use a nontrivial part of the brain energy,
which might in principle be detectable using imaging techniques (Logothetis 2008).

It should be also noted that the above low-cost percentages for synaptic plasticity
agree qualitatively with the marginal role of protein synthesis and actin treadmilling
in the metabolism of the whole brain (Rolfe and Brown 1997; Attwell and Laughlin 2001).
Recent estimates in (Engl and Attwell 2015; Engl et al 2017) suggest however that actin
treadmilling might be much more energy demanding, and overall that nonsignaling
processes in the brain might require similar amounts of energy as those related to
electric signaling (e.g. fast synaptic transmission). However, these estimates were
made on data from brain slices, which may significantly differ from in vivo conditions.
Additionally, they apply to the neurons as a whole, not specifically to synapses, as
in this study. More importantly, Engl and Attwell (2015) and Engl et al (2017) consider
nonsignaling processes, like lipid synthesis, microtubule turnover, and mitochondrial proton
leak, which can be metabolically expensive, but which were not included in the present
analysis for synaptic plasticity, primarily because they seem not to be directly related
to plasticity processes.

The main reasons for the high metabolic efficiency of synaptic plasticity 
is the slowness of molecular processes and small number of molecules in spines related
to learning and memory. For instance, protein synthesis is very slow with a mean time
constant of about 3.7 days (Cohen et al 2013), which is orders of magnitude larger than
the characteristic times associated with short-term synaptic signaling through AMPA and
NMDA receptors (respectively $\sim 5$ and 150 msec), and Na-K-ATPase pump activity
(for pumping out Na$^{+}$ ions) operating on a time scale of a few seconds (Attwell and
Laughlin 2001; Karbowski 2009). On the other hand, faster molecular processes associated
with spine plasticity and related to activated protein phosphorylation, actin treadmilling
and receptor trafficking, operating on a time scale from $0.1-1$ sec to 20-40 sec
(i.e. comparable or even faster to Na-K-ATPase) involve a relatively small number of
molecules that are 2-3 orders of magnitude smaller than the number of
Na$^{+}$ ions that have to be extruded following synaptic transmission.
This high energy efficiency of memory on a molecular level is reminiscent of
energy efficient sparse neural codes on a cellular level (Levy and Baxter 1996; 
Laughlin et al 1998).

\vspace{0.35cm}

\noindent
{\large \bf Limitations of the empirical estimates.}

Every estimation based on incomplete data is only approximate, and this is also the case
for the above results based on molecular data. It seems that the greatest uncertainties
are associated with the protein phosphorylation and plasticity modulation.

The rates of protein phosphorylation can vary by a factor of 10 for the active LTP phase.
The precise values for each PSD protein are not known; at best we have some data for
CaMKII, which fortunately is the most abundant protein in PSD (Sheng and Kim 2011).
Additionally, the metabolic cost of phosphorylation depends on the fraction of proteins
with elevated phosphorylation rates, which in turn depends on the presynaptic stimulation
by Ca$^{2+}$ influx. Based on limited data, we assumed that this fraction is about $10\%$.
However, if it were $30\%$ (under some conditions), then this would increase the overall
cost of synaptic plasticity by a factor 3 to about $12-33\%$ of the synaptic transmission
cost. The maximal possible value for the cost of synaptic plasticity was estimated above
as $15\cdot 10^{6}$ ATP/min, which means that in a hypothetical situation when all PSD
proteins are maximally phosphorylated then the metabolic rate of synaptic plasticity is
about twice of that for synaptic transmission. This value provides an upper bound on the
theoretically possible cost of synaptic plasticity.

The cost of plasticity modulation is hard to compute exactly, as we are uncertain which
pathways are affected and to what extent. We took only one of the possible ways of modulation
(through P2X receptors), and even for this case we do not have numerical values of the
transition rates modulation. The modulation cost was deduced based on the ``end product'',
i.e., the extent to which the synaptic weight can change.

\vspace{0.35cm}

\noindent
{\large \bf Cascade model with phosphorylation cycles reveals metabolic efficiency
     of synaptic memory.}

The main theoretical findings of this study are as follows.
(i) There is a nonzero cost of storing all previous plasticity events, i.e. prior memory,
which is kept in baseline entropy production rate EPR$_{0}$ (Figs. 2 and 3).
(ii) In most cases, the metabolic cost of a new learning and its memory trace increases
proportionally with memory duration, such that their ratio is essentially constant (Figs. 5-7).
(iii) A different trend is observed for the two key parameters, slowing down factor $z$ and
ATP driven transition amplitude $a_{0}$, for which a memory lifetime per expanded energy grows
for longer memories (Figs. 8 and 9). (iv) Memory trace decays with time as a power law, while
the associated with it metabolic rate decays to its baseline much faster, exponentially,
which leads to the dynamic decoupling of memory trace and its metabolism (Fig. 3).
The likely reason for the fast decay of the metabolic rate is that EPR depends nonlinearly
on the probabilities and transition rates.
(v) The direct consequence of (iv) is that maintenance of a new memory takes only
a small fraction of the energy required for the baseline synaptic state, i.e. the ratio
$\Delta E/E_{0} \ll 1$ (Figs. 6-9). Additionally, the huge majority of energy is expanded
during the learning phase (synaptic stimulation), and the memory phase (after the stimulation)
is relatively costless (Fig. 3).

Points (ii) and (iii) indicate that while there is some cost to learning and storing
a new memory, it can be minimized if the right parameters are appropriately tuned. Points
(iv) and (v) indicate that memory trace and its metabolic energy consumption are governed
by different dynamical time scales and are essentially independent for long times,
which leads to a relatively low cost of a new memory. Both effects are more pronounced
if the slowing down factor $z$ is large, which indicates that the diversity of time
scales associated with biochemical synaptic cascades increases not only the memory
lifetime (Fusi et al 2005; Benna and Fusi 2016), but also enhances its metabolic efficiency. 
Even bigger metabolic efficiency is obtained if the ATP driven phosphorylation
amplitude $a_{0}$ is made larger (Fig. 9). Surprisingly, in this case it is possible
to have longer memories for less energy if $a_{0}$ and $z$ are sufficiently large.

How can we explain the effects of $a_{0}$ and $z$ on the energetic efficiency of
memory? They can be explained by noting that increasing $z$ causes decreasing
the speed of all molecular processes that in turn consume less energy per time unit
(Fig. 2, middle panel). Additionally, the resting metabolic rate EPR$_{0}$
depends non-monotonically on $a_{0}$ for $z > 0$, and generally EPR$_{0}$ decreases
with increasing $a_{0}$ if $a_{0}$ is large (Fig. 2, bottom panel). In this case
the ratio $T_{m}/E$ has a minimum for some $a_{0}$ if $z > 0$ (Fig. 9), but then
$T_{m}/E$ generally steady increases for large $a_{0}$, suggesting an enhanced
metabolic efficiency of synaptic memory in this regime.

It should be noted that the energy cost of a new learning and its memory trace
found here is a theoretical minimum, because the cascade model considers only protein
phosphorylation. Inclusion of the other energy consuming molecular processes would be
much more complicated and would require a much more complex model that goes far beyond
simple Markov models analyzed within a Master equation approach.

\vspace{0.35cm}

\noindent
{\large \bf Model of synaptic plasticity as phosphorylation cascades.} 

The cascade model of synaptic plasticity presented in this study differs from
the previous cascade models (Fusi et al 2005; Leibold and Kempter 2008; Barrett
et al 2009; Benna and Fusi 2016). First, the transitions between synaptic states
are bidirectional here, as opposed to the majority of the previous models
(see however, Benna and Fusi 2016, as an exception). Second, the present model
contains many local cyclic motifs, corresponding to ATP driven protein
phosphorylation, which are absent in the previous models. Strictly speaking,
there are some loops in the topology of the previous models but they are
unidirectional and non local, and hence it is difficult to interpret their
physical meaning. Third, and most importantly, the current study ask a fundamentally
different question, namely the energy cost and efficiency of memory on a spine level.
It should be stressed that, without bidirectional cyclic motifs, most of the previous
cascade models are inappropriate for studying metabolic constraints on memory,
since they are thermodynamically inconsistent and produce singular entropy
production rate (metabolic rate), as explained in this study (Qian 2006, 2007).

These differences generate also qualitative differences in some results. For instance,
the signal to noise memory trace SNR decays here for long times as $\sim t^{-4/3}$ (Fig. 3),
which is much faster than in Benna and Fusi (2016), where it decays as $\sim t^{-1/2}$.
The primary reason for this is that the Benna and Fusi (2016) model is optimized,
while the current model is not. Another difference is that here the memory
lifetime always saturates (Figs. 5-9), also as a function of the number of basic
states $n$ (Fig. 6). In contrast, in the Benna and Fusi (2016) model the memory
duration increases exponentially with synaptic complexity, which may be equivalent
to $n$. The likely explanation for this difference is the presence of many local
cycles in the structure of the current model, which provide additional pathways for
faster relaxation to baseline conditions.

\vspace{0.35cm}

\noindent
{\large \bf Limitations of the present model.} 

The model presented in this study is an obvious simplification of a much more
complicated web of molecular interactions in a typical spine. However, the goal
here was not to model the spine in its full complexity, but rather to identify key
parameters related to energy consumption during memory storage that are most sensitive
in terms of metabolic efficiency, and this could be best done for simple models.
Nevertheless, the current model can be modified and extended by including specific
molecular details, such as those considered in several previous studies
(Miller et al 2005; Hayer and Bhalla 2005; Graupner and Brunel 2007; Bhalla 2011;
Antunes and Schutter 2012; Smolen et al 2012; Kim et al 2013).

\newpage

\noindent
{\large \bf Implications of energy efficient memory storage.} 

The empirical estimates as well as the theoretical analysis performed here imply
that molecular storing of memory can be relatively cheap under some conditions
($\le 11\%$ of the synaptic transmission cost).
This result can have important implications for building energy efficient artificial
silicon systems that mimic brain function by storing and processing information
(Esser et al 2016; Sun et al 2018). The key here is to have bidirectional loops in
the topology of subsynaptic biochemical pathways with appropriately tuned transition
rates and multiple time constants, not only to prevent catastrophic forgetting in
neural networks (Kirkpatrick et al 2017) but also to spend small amounts
of energy on long-term synaptic computations.

Another potential implication of the current results is for biomedical research
related to neurodevelopmental disorders. There are experimental data showing a
close relationship between phosphorylation signaling in PSD and diseases such
as schizophrenia and autism (Li et al 2016). Interestingly, many genes encoding PSD
overlap with mutated genes responsible for schizophrenia and for autistic phenotype
(De Rubeis et al 2014; Iossifov et al 2014; Fromer et al 2014; Kaizuka and Takumi
2018). In the light of the results obtained in this study, it is not difficult to
understand this close relationship given the easiness with which one can alter memory
lifetime and its energetic efficiency by manipulating transition rates related to
ATP driven protein phosphorylation. Moreover, there are many empirical studies
showing that increased glucose metabolism can enhance memory in mammals, which is
known as ``glucose memory facilitation effect'', in a dose-dependent fashion
(Gold 2005; Smith et al 2011). The latter means that some glucose levels
can facilitate memory, while others can be neutral or even detrimental for memory.
Using our model, this phenomenon can be understood by noting that ATP that drives
protein phosphorylation is generated directly by glucose (Rolfe and Brown 1997).
This means that glucose metabolic rates can affect ATP rates used for powering
spine PSD proteins and downstream molecular processes related to synaptic information
storing, e.g. as shown in Fig. (9).

\newpage

{\large \bf METHODS}

\vspace{0.35cm}

\noindent
{\large \bf Cascade model of synaptic plasticity with bidirectional cycles.}

We model biochemical processes underlying synaptic plasticity in a dendritic spine
as transitions between discrete synaptic states (Montgomery and Madison 2004), which
describe various levels of protein activation in spine PSD (Fig. 1).
The model of plasticity considered here is a generalization of previous 
models of cascade synaptic plasticity (Fusi et al 2005) and is treated as a Master
Equation system (Schnakenberg 1976). The main modification in the current approach
is the addition of closed loops or cycles with bidirectional transitions between
different states, corresponding to protein phosphorylation (Qian 2006).

A synapse can be in one of many ``down'' and ``up'' states corresponding to biochemical 
process associated with LTD and LTP, respectively (Fig. 1). There are $n$ basic up states
and $n$ basic down states, whose probabilities of occupancy are given by $p_{k}^{+}$
and $p_{k}^{-}$, respectively for $k=1,...,n$. There are two types of transitions between
neighboring $p^{\pm}_{k}$ states: direct spontaneous transitions driven by thermal fluctuations 
(with transition rates $\alpha_{k}^{\pm}$ and $\epsilon_{1}\alpha_{k}^{\pm}$), and indirect 
transitions that contain ATP driven reactions (with transition rates $a^{\pm}_{k}$ and
$\epsilon_{3} b$) and non-ATP reactions (with transition rates $\beta^{\pm}_{k}$ and
$\epsilon_{2}\beta^{\pm}_{k}$). The direct transitions are rare, while the indirect ones
can be massive if the synapse is electrically stimulated in a way that induces plasticity
mechanisms (LTP and LTD), and elevation of local ATP concentration. Additionally,
the ATP related transitions require intermediate states with probabilities of occupancy
$q^{\pm}_{k}$, which describe metastable states with proteins ready for phosphorylation
(for details see, e.g., Qian 2007).
Each loop or cyclic motif represented by transitions
$p^{\pm}_{k} \mapsto q_{k}^{\pm} \mapsto p^{\pm}_{k+1} \mapsto p_{k}^{\pm}$, can be thought
as a cascading phosphorylation of various downstream proteins in spine PSD.

It is also assumed that the dynamics of downstream cascades are progressively slower,
similar to the previous models (Fusi et al 2005; Benna and Fusi 2016).
Mathematically, this is implemented by a prefactor $g_{k}= \exp(-zk)$,
which rescales all the transition rates and gets smaller for deeper
states with higher index $k$, where $z$ is the slowing down factor.

The dynamics of state probabilities are given by

\begin{eqnarray}
\dot{p}^{\pm}_{k} = g_{k-1}(\epsilon_{1}\alpha_{k-1}^{\pm}p^{\pm}_{k-1}+a^{\pm}_{k-1} q^{\pm}_{k-1}) 
+ g_{k}(\alpha_{k}^{\pm}p^{\pm}_{k+1}+\beta^{\pm}_{k} q^{\pm}_{k})      \\  \nonumber 
- \left[g_{k-1}(\epsilon_{3}b_{0}+\alpha^{\pm}_{k-1}) 
+ g_{k}(\epsilon_{1}\alpha_{k}^{\pm}+\epsilon_{2}\beta^{\pm}_{k})\right]p^{\pm}_{k} 
\end{eqnarray}\\
for $2 \le k \le n-1$, where the dot denotes the time derivative.
For $k=n$ we have

\begin{eqnarray}
  \dot{p}^{\pm}_{n} =  g_{n-1}\left( a^{\pm}_{n-1}q^{\pm}_{n-1}
  + \epsilon_{1}\alpha^{\pm}_{n-1} p^{\pm}_{n-1} 
- (\epsilon_{3}b_{0}+\alpha^{\pm}_{n-1})p^{\pm}_{n} \right),
\end{eqnarray}\\
and for $k=1$ we have

\begin{eqnarray}
\dot{p}^{+}_{1} = \epsilon_{1}\alpha_{0}p^{-}_{1} + a_{0} q_{0} +
g_{1}(\alpha_{1}^{+}p^{+}_{2}+\beta^{+}_{1} q^{+}_{1})      \\  \nonumber 
- \left[g_{1}(\epsilon_{1}\alpha_{1}^{+} +\epsilon_{2}\beta^{+}_{1}) + \alpha_{0}
  + \epsilon_{3}b_{0}\right]p^{+}_{1} 
\end{eqnarray}\\
and
\begin{eqnarray}
\dot{p}^{-}_{1} = \alpha_{0}p^{+}_{1} + \beta_{0}q_{0} +
g_{1}(\alpha_{1}^{-}p^{-}_{2}+\beta^{-}_{1}q^{-}_{1})      \\  \nonumber 
- \left[g_{1}(\epsilon_{1}\alpha_{1}^{-} +\epsilon_{2}\beta^{-}_{1})
  + \epsilon_{1}\alpha_{0} + \epsilon_{2}\beta_{0}\right]p^{-}_{1}.
\end{eqnarray}

The dynamics of the intermediate states involved in protein phosphorylation
are given by

\begin{eqnarray}
  \dot{q}^{\pm}_{k} =  g_{k}\left(\epsilon_{2}\beta^{\pm}_{k} p^{\pm}_{k}
  + \epsilon_{3}b_{0} p^{\pm}_{k+1} 
- (a^{\pm}_{k}+\beta^{\pm}_{k})q^{\pm}_{k} \right),
\end{eqnarray}\\
for $1 \le k \le n$, and

\begin{eqnarray}
\dot{q}_{0} = \epsilon_{2}\beta_{0} p^{-}_{1} + \epsilon_{3}b_{0} p^{+}_{1} 
- (a_{0}+\beta_{0})q_{0}.
\end{eqnarray}\\
The transition rates $\alpha_{k}^{\pm}, \beta_{k}^{\pm}, a_{k}^{\pm}$ 
are heterogeneous across different cycles and given by
$\alpha_{k}^{\pm}= \alpha_{0}(1+\sigma\eta^{\pm}_{k})$,
$\beta_{k}^{\pm}= \beta_{0}(1+\sigma\eta^{\pm}_{k})$, and
$a_{k}^{\pm}= a_{0}(1+\sigma\eta^{\pm}_{k})$, where $\eta^{\pm}_{k}$ are
random variables uniformly distributed between $-1$ and 1, and $\sigma$
is a measure of heterogeneity ($0 \le \sigma < 1$). The parameters
$\alpha_{0}, \beta_{0}$, and $a_{0}$ are the amplitudes of the above transition rates.
When the synapses are stimulated, only the ATP-driven rates $a_{k}^{\pm}, a_{0}$
are time dependent (see below).

Let us consider two examples of 3 state models: one with a ladder structure
(Fig. 1A), and another with a loop structure (Fig. 1B). For the ladder
structure (Fig. 1A) we have the dynamics of state probabilities $p_{1}$, $p_{2}, p_{3}$
as

\begin{eqnarray}
\dot{p}_{1} = J_{12},   \nonumber \\ 
\dot{p}_{2} = J_{21} - J_{32},   \nonumber \\ 
\dot{p}_{3} = J_{32},
\end{eqnarray}\\
where the probability fluxes are defined as:
$J_{12}= w_{12}p_{2}-w_{21}p_{1}$  (flux from state 2 to state 1),
$J_{32}= w_{32}p_{2}-w_{23}p_{3}$  (flux from state 2 to state 3),
$J_{21}= -J_{12}$  (flux from state 1 to state 2). At the steady-state
$(\dot{p}_{1}= \dot{p}_{2}= \dot{p}_{3}= 0)$, we obtain $w_{12}p_{2}= w_{21}p_{1}$,
and $w_{32}p_{2}= w_{23}p_{3}$. From this it follows that all steady-state
fluxes $J_{12}= J_{32}= 0$, which is known as the condition of the detailed
balance (Lebowitz and Spohn 1999; Mehta and Schwab 2012).

For the second example with a loop (Fig. 1B), the dynamics of
state probabilities $p_{+}$, $p_{-}, q_{0}$ read:

\begin{eqnarray}
\dot{p}_{+} = J_{+0} + J_{+-},    \nonumber \\ 
\dot{p}_{-} = J_{-0} - J_{+-},      \nonumber \\ 
\dot{q}_{0} = -J_{+0} - J_{-0},
\end{eqnarray}\\
where the probability fluxes are defined as:
$J_{+0}= aq_{0} - \epsilon_{3}bp_{+}$ (flux from state 0 to $+$),
$J_{+-}= \epsilon_{1}\alpha p_{-} - \alpha p_{+}$ (flux from state $-$ to $+$),
$J_{-0}= \beta q_{0} - \epsilon_{2}\beta p_{-}$ (flux from state 0 to $-$),
and the opposite fluxes are $J_{0+}= -J_{+0}$, $J_{-+}= -J_{+-}$, and $J_{0-}= -J_{-0}$.
For this cyclic motif, we can find steady-state
values of the probabilities as:

\begin{eqnarray}
  p_{+}= Z^{-1}\left[\epsilon_{1}\alpha(a+\beta) + \epsilon_{2}a\beta\right], \nonumber  \\
  p_{-}= Z^{-1}\left[\alpha(a+\beta) + \epsilon_{3}b\beta\right],  \nonumber  \\
  q_{0}= Z^{-1}\left[\epsilon_{2}\alpha\beta + \epsilon_{3}(\epsilon_{1}\alpha
    + \epsilon_{2}\beta)\right],
\end{eqnarray}\\
where $Z$ is given by

\begin{eqnarray}
Z= \alpha[(1+\epsilon_{1}+\epsilon_{2})\beta + (1+\epsilon_{1})a 
+ \epsilon_{1}\epsilon_{3}b] + \beta[\epsilon_{2}a+\epsilon_{3}(1+\epsilon_{2})b].
\end{eqnarray}\\
  
At the steady-state the above fluxes must balance each other, i.e.,
$J_{+0}= - J_{+-}= -J_{-0}\equiv J$, where the emerging flux 
$J= \alpha\beta Z^{-1}(\epsilon_{1}\epsilon_{3}b - \epsilon_{2}a)$.
Note that the flux $J$ is generally nonzero, which is a signature of a nonequilibrium
steady-state, denoted as NESS (Lebowitz and Spohn 1999; Bustamante et al 2005;
Van den Broeck and Esposito 2015).
It vanishes only if the phosphorylation and dephosphorylation rates
($a$ and $\epsilon_{3}b$) are both zero (cyclic motif is destroyed), or for the special
case of the so-called detailed balance when $\epsilon_{1}\epsilon_{3}b = \epsilon_{2}a$.
The latter two situations correspond to thermodynamic equilibrium when neither energy
nor material is exchanged with the environment (``thermodynamic death'', e.g.,
Nicolis and Prigogine 1977).

\vspace{0.35cm}

\noindent
{\large \bf Memory trace and signal to noise ratio.}

We consider $N_{s}$ independent synapses for which we first determine their non-equilibrium
steady-state NESS. This is done by starting from uniform initial conditions for the
state probabilities $p_{k}^{\pm}, q_{k}^{\pm}$ and allowing them to relax to the baseline
state denoted as $p_{k,\infty}^{\pm}, q_{k,\infty}^{\pm}$. A next phase is a brief stimulation
of synapses from their baseline and observation of the associated memory lifetime
of this event. During the stimulation and subsequent memory decay, the synapses are
divided into two populations: the fraction $f$ of synapses undergoes LTP process,
and the remaining $1-f$ synapses perform LTD process. The LTP synapses are stimulated
by a pulse in the ATP-driven transitions $a_{0}, a_{k}^{+}$, while the LTD synapses are
activated by a pulse in the ATP-driven transitions $a_{k}^{-}$, with the explicit
time dependences given by $a_{k}^{\pm}(t)= a_{k}^{\pm}[1 + A_{\pm}\exp(-t/\tau)]$
for $k > 1$, and $a_{0}(t)= a_{0}[1 + A_{+}\exp(-t/\tau)]$,
where $\tau$ is the characteristic time of stimulation, and $t$ is the time counted
from the onset of stimulation. We assume that $A_{+} > A_{-}$, which reflects an
experimental fact that LTP (LTD) is induced by high (low) frequency stimulation.

We assume that all down states (including $q_{0}$) have the same synaptic efficacy 
(weight) $w$, and all up states have the same efficacy $2w$, where $w$ is the synaptic 
conductance. (Its value is irrelevant for the results of this study.)
This binary choice is consistent with neurophysiological data (Petersen et al 1998;
O'Connor et al 2005). Thus, the probability that a randomly chosen synapse undergoes
LTD and has weight $w$ is $(1-f)(p^{-}_{LTD} + q^{-}_{LTD})$, and the probability that it
has weight $2w$ is $(1-f)(p^{+}_{LTD} + q^{+}_{LTD})$, where
$p_{LTD}^{\pm} = \sum_{k} p_{k,LTD}^{\pm}$, and
$q_{LTD}^{\pm} = \sum_{k} q_{k,LTD}^{\pm}$.
Similarly, the probability that a randomly selected synapse undergoes LTP and has
weight $w$ is $f(p^{-}_{LTP} + q^{-}_{LTP})$, and that it has weight $2w$ is
$f(p^{+}_{LTP} + q^{+}_{LTP})$, where
$p_{LTP}^{\pm} = \sum_{k} p_{k,LTP}^{\pm}$, and
$q_{LTP}^{\pm} = \sum_{k} q_{k,LTP}^{\pm} + q_{0}$.
From this it follows that the average synaptic weight $\langle V\rangle$ for the
whole synaptic population is given by

\begin{eqnarray}
 \langle V\rangle= w\left[ (1-f)(p^{-}_{LTD} + q^{-}_{LTD}) + f(p^{-}_{LTP} + q^{-}_{LTP}) \right]
  \nonumber \\
 + 2w\left[ (1-f)(p^{+}_{LTD} + q^{+}_{LTD}) + f(p^{+}_{LTP} + q^{+}_{LTP}) \right].
\end{eqnarray}\\
Consequently, the variance in a population synaptic weight, i.e.
$\langle V^{2}\rangle - \langle V\rangle^{2}$, is

\begin{eqnarray}
 \langle V^{2}\rangle - \langle V\rangle^{2} =
  w^{2}\left[ (1-f)(p^{-}_{LTD} + q^{-}_{LTD}) + f(p^{-}_{LTP} + q^{-}_{LTP}) \right]
  \nonumber \\
 \times \left[ (1-f)(p^{+}_{LTD} + q^{+}_{LTD}) + f(p^{+}_{LTP} + q^{+}_{LTP}) \right].
\end{eqnarray}\\

We define a synaptic memory trace as a deviation of the average synaptic weight
$\langle V\rangle$ from its baseline value $\langle V\rangle_{b}$, and normalized
by a standard deviation in $V$ (similar to Fusi et al 2005).
This is equivalent to the definition of signal to noise ratio SNR at time $t$:

\begin{eqnarray}
  \mbox{SNR}(t)= \sqrt{N_{s}} \frac{(\langle V\rangle - \langle V\rangle_{b})}
  {\sqrt{\langle V^{2}\rangle  - \langle V\rangle^{2}}},
\end{eqnarray}\\
where the prefactor $\sqrt{N_{s}}$ comes from summing contributions from all the
synapses in a small cortical region. Note that after a synaptic stimulation both, the
signal $\langle V\rangle$ and the variance in the denominator, are time dependent.
Note also that SNR does not depend on the value of synaptic weight $w$ (it cancels out).
It is assumed that when SNR drops below a value of 1,
the memory trace becomes undetectable and this time determines the
memory lifetime $T_{m}$. The general results and conclusions are independent of the
precise choice of this threshold.

\vspace{0.35cm}

\noindent
{\large \bf Entropy production as a synaptic metabolic rate.}

Metabolic rate (or the rate of dissipated energy) associated with cascading biochemical
processes in a synapse is associated with entropy production rate EPR,
which is defined as (Schnakenberg 1976; Lebowitz and Spohn 1999; Van den Broeck
and Esposito 2015)

\begin{eqnarray}
\mbox{EPR}= \frac{1}{2}kT \sum_{i,j} (w_{ij}P_{j}-w_{ji}P_{i})
\ln\frac{w_{ij}P_{j}}{w_{ji}P_{i}}
\end{eqnarray}\\
where $k$ is the Boltzmann constant and $T$ the is brain temperature, 
$w_{ij}$ are the transition rates between states $j$ and $i$, and $P_{i}$
is the probability of the state $i$ occupancy. The above EPR should be understood as an
energy rate per a single biochemical cascade; in the case of many cascades, the result
should be multiplied by the number of pathways. Note that when the transition 
between two given states is unidirectional, then one of transition rates in
a pair (either $w_{ij}$ or $w_{ji}$) must vanish. This means that a corresponding 
logarithm in the sum must diverge to infinity, which implies a divergent metabolic
rate. This is clearly not a realistic description of the energetics of any 
biological system, which suggests that one must always keep all the transition rates
as bidirectional (regardless of how small they are).
Unfortunately, this important fact was overlooked in early models of cascade
plasticity, where many transitions were chosen as unidirectional (Fusi et al 2005;
Leibold and Kempter 2008; Barrett et al 2009), and only recently it was realized that
bidirectionality is important for a memory lifetime duration (Benna and Fusi 2016).

In a particular case of a simple 3 state model with a ladder structure
(Fig. 1A), the entropy production rate at steady-state is zero. This is due to
the detailed balance condition $w_{12}p_{2}= w_{21}p_{1}$ and $w_{32}p_{2}= w_{23}p_{3}$,
which implies that all the terms in Eq. (15) are zero. This situation corresponds
to a vanishing flux, which means that our 3 state system at steady state is in
thermal equilibrium with its environment.

On the contrary, for the 3 state model with a loop (Fig. 1B), the detailed balance
is broken, and there is a circulating flux even at steady-state, which leads to
a non zero EPR given by Eq. (1).

For our plasticity model with $2n-1$ basic cyclic motifs (Fig. 1C), EPR has five major
contributions: two caused by direct transitions within down states (EPR$^{-}_{pp}$)
and within up states (EPR$^{+}_{pp}$), two contributions caused by indirect transitions
involving $p$ and $q$ states either for down (EPR$^{-}_{pq}$) or for up states (EPR$^{+}_{pq}$), 
and one contribution related to transitions within the basic loop 
$p^{-}_{1} \mapsto p^{+}_{1} \mapsto q_{0} \mapsto p^{-}_{1}$ (EPR$_{ppq}$).
Thus, the EPR associated with synaptic plasticity takes the form

\begin{eqnarray}
\mbox{EPR}= \mbox{EPR}^{-}_{pp} + \mbox{EPR}^{+}_{pp} +
\mbox{EPR}^{-}_{pq} + \mbox{EPR}^{+}_{pq} + \mbox{EPR}_{ppq},
\end{eqnarray}\\
where the appropriate contributions read

\begin{eqnarray}
\mbox{EPR}^{\pm}_{pp}=
\sum_{k=1}^{n-1} g_{k}\alpha_{k}^{\pm}(\epsilon_{1}p^{\pm}_{k}-p^{\pm}_{k+1})
\ln\left(\frac{\epsilon_{1}p^{\pm}_{k}}{p^{\pm}_{k+1}}\right),
\end{eqnarray}\\

\begin{eqnarray}
\mbox{EPR}^{\pm}_{pq}=
\sum_{k=1}^{n-1} g_{k} \left[ \beta_{k}^{\pm}
(\epsilon_{2} p^{\pm}_{k}-q^{\pm}_{k})
\ln\left(\frac{\epsilon_{2}p^{\pm}_{k}}{q^{\pm}_{k}}\right) +
(\epsilon_{3}b_{0}p^{\pm}_{k+1}-a^{\pm}_{k}q^{\pm}_{k})
\ln\left(\frac{\epsilon_{3}b_{0}p^{\pm}_{k+1}}{a^{\pm}_{k}q^{\pm}_{k}}\right)
\right],
\end{eqnarray}\\
and the basic loop contribution is

\begin{eqnarray}
 \mbox{EPR}_{ppq}= \alpha_{0}(p^{+}_{1}-\epsilon_{1}p^{-}_{1})
 \ln\left(\frac{p^{+}_{1}}{\epsilon_{1}p^{-}_{1}}\right)  +
 \beta_{0}(q_{0}-\epsilon_{2}p^{-}_{1})\ln\left(\frac{q_{0}}{\epsilon_{2}p^{-}_{1}}\right)
 \\  \nonumber
 + (a_{0}q_{0}-\epsilon_{3}b_{0}p^{+}_{1})
 \ln\left(\frac{a_{0}q_{0}}{\epsilon_{3}b_{0}p^{+}_{1}}\right).
\end{eqnarray}\\

Energy used for synaptic stimulation and subsequent recovery to NESS state, which
is the energy needed to keep a memory trace above the threshold is defined as

\begin{eqnarray}
E= \int_{0}^{T_{m}} dt \; \mbox{EPR}(t),
\end{eqnarray}\\
where $t=0$ relates to the moment of stimulation. The relative energy used for maintaining
memory is defined as the ratio $(E-E_{0})/E_{0}$, where $E_{0}= \mbox{EPR}_{0}T_{m}$ is
the baseline energy used during the time interval of duration $T_{m}$, and EPR$_{0}$ is
the baseline entropy production rate.

\vspace{0.35cm}

\noindent
{\large \bf Parameters used in the model.}

The following default values of the parameters were used. For the transition rates:
$a_{0}= 0.4$ min$^{-1}$, $b_{0}= 0.2$ min$^{-1}$ (Molden et al 2014), $\alpha_{0}= 0.05$
min$^{-1}$, $\beta_{0}= 20.0$ min$^{-1}$ (Miller et al 2005), $\epsilon_{1}= 0.001$,
$\epsilon_{2}= 0.05$, $\epsilon_{3}= 0.0001$ (there is high asymmetry between reaction
rates for protein phosphorylation and dephosphorylation, see Qian 2006 and 2007),
$\sigma= 0.25$. Note that the direct spontaneous transitions for protein activation
induced by thermal fluctuations are much weaker than intermediate transitions
associated with ATP hydrolysis. Default values for synaptic stimulation:
$A_{+}= 50$, $A_{-}= 10$, and $\tau= 10$ min. Other values: number of synapses
$N_{s}= 10^{7}$ (typical number in a cortical column with $10^{3}$ neurons),
number of states for up and down configurations $n= 5$, fraction of potentiated
synapses $f= 0.5$, and the slowing-down rate $z=0.8$.
All the figures are made for these default values unless indicated otherwise.

\vspace{1.8cm}

\noindent{\large \bf Acknowledgments}

The work was supported by the Polish National Science Centre 
(NCN) grant no. 2015/17/B/NZ4/02600.

\vspace{1.8cm}

\noindent{\bf Declaration of interests}

The author declares no competing interests.

\newpage

\vspace{1.5cm}

\noindent{\bf\large  References} \\
Abraham WC, Bear MF (1996) Metaplasticity: the plasticity of synaptic plasticity.
{\it Trends Neurosci.} {\bf 19}: 126-130.  \\
Aiello LC, Wheeler P (1995) The expensive-tissue hypothesis: The brain
and the digestive-system in human and primate evolution. {\it Curr.
Anthropology} {\bf 36}: 199-221.   \\
Antunes G, Schutter ED (2012) A stochastic signaling network mediates
the probabilistic induction of cerebellar long-term depression.
{\it J. Neurosci.} {\bf 32}: 9288-9300.  \\
Attwell D, Laughlin SB (2001) An energy budget for signaling in the 
gray matter of the brain. {\it J. Cereb. Blood Flow Metabol.} 
{\bf 21}: 1133-1145.  \\
Barrett AB, Billings GO, Morris RGM, van Rossum MCW (2009) State based model
of long-term potentiation and synaptic tagging and capture.
{\it PLoS Comput. Biol.} {\bf 5}: e1000259.   \\
Bayes A, Collins MO, Croning MD, van de Lagemaat LN, Choudhary JS, Grant SG
(2012) Comparative study of human and mouse postsynaptic proteomes finds
high compositional conservation and abundance differences for key synaptic
proteins. {\it PLoS ONE} {\bf 7}: e46683.  \\
Bean BP, Williams CA, Ceelen PW (1990) ATP-activated channels in rat and bullfrog
sensory neurons: current-voltage relation and single-channel behavior.
{\it J. Neurosci.}  {\bf 10}: 11-19.  \\
Benna MK, Fusi S (2016) Computational principles of synaptic memory
consolidation. {\it Nature Neurosci.} {\bf 19}: 1697-1706.    \\
Bhalla US (2011) Trafficking motifs as the basis for two-compartment
signaling systems to form multiple stable states. {\it Biophys. J.}
{\bf 101}: 21-32.    \\
Bhalla US, Iyengar R (1999) Emergent properties of networks of biological
signaling pathways. {\it Science} {\bf 283}: 381-387.   \\
 Bhalla US (2014) Molecular computation in neurons: a modeling perspective.
  {\it Curr. Opin. Neurobiol.} {\bf 25}: 31-37.   \\
Bosch M, Castro J, Saneyoshi T, Matsuno H, Sur M, Hayashi Y (2014) Structural
and molecular remodeling of dendritic spine substructures during long-term
potentiation. {\it Neuron} {\bf 82}: 444-459.    \\
Borgdorff AJ, Choquet D (2002) Regulation of AMPA receptor lateral movements.
  {\it Nature}  {\bf 417}: 649-653.     \\
Bradshaw JM, Hudmon A, Schulman H (2002) Chemical quenched flow kinetic
studies indicate an intraholoenzyme autophosphorylation mechanism for
Ca$^{2+}$/Calmodulin-dependent protein kinase II.
{\it J. Biol. Chem.} {\bf 277}: 20991-20998.  \\
 Bustamante C, Liphardt J, Ritort F (2005) The nonequilibrium thermodynamics of
  small systems. {\it Phys. Today} {\bf 58}: 43.  \\
Chaudhuri R, Fiete I (2016) Computational principles of memory.
{\it Nature Neurosci.} {\bf 19}: 394-403.   \\
  Chen X, Vinade L, Leapman RD, Petersen JD, Nakagawa T, Phillips TM, Sheng M,
  Reese TS (2005) Mass of the postsynaptic density and enumeration of three
  key molecules. {\it Proc. Natl. Acad. Sci. USA} {\bf 102}: 11551-11556.   \\
Choquet D, Triller A (2013) The dynamic synapse. {\it Neuron} {\bf 80}: 691-703.  \\
Cingolani LA, Goda Y (2008) Actin in action: the interplay between the actin
cytoskeleton and synaptic efficacy. {\it Nat. Rev. Neurosci.} {\bf 9}: 344-356. \\
Clarke DD, Sokoloff L (1994) In Siegel GJ et al (eds), {\it Basic Neurochemistry},
pp. 645-680. Raven: New York, NY.   \\
 Coba MP, Pocklington AJ, Collins MO, Kopanitsa MV, Uren RT, Swamy S, Croning MD,
  Choudhary JS, Grant SG (2009) Neurotransmitters drive combinatorial multistate
  postsynaptic density networks. {\it Sci. Signal.} {\bf 2}: ra19.  \\
Cohen LD, Zuchman R, Sorokina O, Muller A, Dieterich DC, Armstrong JD, et al.
(2013) Metabolic turnover of synaptic proteins: kinetics, interdependencies
and implications for synaptic maintenance. {\it PLoS ONE} {\bf 8}: e63191.  \\
Colbran RJ (1993) Inactivation of Ca$^{2+}$/Calmodulin-dependent protein kinase
II by basal autophosphorylation. {\it J. Biol. Chem.} {\bf 268}: 7163-7170.  \\
  Collins MO, Yu L, Coba MP, Husi H, Campuzano I, Blackstock WP, Choudhary JS,
  Grant SCN (2005) Proteomic analysis of in vivo phosphorylated synaptic proteins.
  {\it J. Biol. Chem.} {\bf 280}: 5972-5982.    \\
De Koninck P, Schulman H (1998) Sensitivity of CaM kinase II to the frequency
 of Ca$^{2+}$ oscillations. {\it Science}  {\bf 279}: 227-230.  \\ 
  De Rubeis S, He X, Goldberg AP, Poultney CS, Samocha K, et al (2014) Synaptic,
  transcriptional and chromatin genes disrupted in autism. {\it Nature}  {\bf 515}:
  209-215.   \\
  Devineni N, Minamide LS, Niu M, Safer D, Verma R, Bamburg JR, Nachmias VT
  (1999) A quantitative analysis of G-actin binding proteins and the G-actin
  pool in developing chick brain. {\it Brain Res.} {\bf 823}: 129-140.   \\
Ehlers MD (2000) Reinsertion or degradation of AMPA receptors determined by
activity-dependent endocytic sorting. {\it Neuron}  {\bf 28}: 511-525.  \\
Engl E, Attwell D (2015) Non-signalling energy use in the brain.
{\it J. Physiol.} {\bf 593}: 3417-3429.   \\
Engl E, Jolivet R, Hall CN, Attwell D (2017) Non-signalling energy use in the
developing rat brain. {\it J. Cereb. Blood Flow Metab.} {\bf 37}: 951-966.  \\
Esser SK, Merolla PA, Arthur JV, Cassidy AS, Appuswamy R et al (2016) Convolutional
networks for fast, energy-efficient neuromorphic computing.
{\it Proc. Natl. Acad. Sci. USA}  {\bf 113}: 11441-11446.  \\
Fanselow EE, Nicolelis MAL (1999) Behavioral modulation of tactile responses in the
rat somatosensory system. {\it J. Neurosci.} {\bf 19}: 7603-7616.  \\
Francois-Martin C, Rothman JE, Pincet F (2017) Low energy cost for optimal speed and control
of membrane fusion. {\it Proc. Natl. Acad. Sci. USA} {\bf 114}: 1238-1241.  \\
Fromer M, Pocklington AJ, Kavanagh DH, Williams HJ, Dwyer S, et al (2014)
De novo mutations in schizophrenia implicate synaptic networks.
{\it Nature}  {\bf 506}: 179-184.  \\
Fusi S, Drew PJ, Abbott LF (2005) Cascade models of synaptically stored memories.
{\it Neuron} {\bf 45}: 599-611.    \\
Gaertner TR, Kolodziej SJ, Wang D, Kobayashi R, et al (2004) Comparative analyses
of the three-dimensional structures and enzymatic properties of $\alpha, \beta, \gamma$
and $\delta$ isoforms of Ca$^{2+}$-Calmodulin-dependent protein kinase II.
{\it J. Biol. Chem.} {\bf 279}: 12484-12494.  \\
Gold PE (2005) Glucose and age-related changes in memory.
{\it Neurobiol. Aging} {\bf 26S}: S60-S64.  \\
Gordon GRJ, Baimoukhametova DV, Hewitt SA, Kosala WRA, Rajapaksha JS, Fisher TE,
Bains J (2005) Norepinephrine triggers release of glial ATP to increase
postsynaptic efficacy. {\it Nat. Neurosci.} {\bf 8}: 1078-1086.  \\
Grafmuller A, Shillcock J, Lipowsky R (2009) The fusion of membranes and vesicles:
 pathway and energy barriers from dissipative particle dynamics. {\it Biophys. J}
  {\bf 96}: 2658-2675.  \\
Graupner M, Brunel N (2007) STDP in a bistable synapse model based on CaMKII switch:
dependence on the number of enzyme molecules and protein turnover.
{\it PLoS Comput. Biol.} {\bf 3}: e221.   \\
Gumbart J, Chipot C, Schulten K (2011) Free-energy cost for translocon-assisted
 insertion of membrane proteins. {\it Proc. Natl. Acad. Sci. USA} {\bf 108}: 3596-3601.
Harris JJ, Jolivet R, Attwell D (2012) Synaptic energy use and supply.
{\it Neuron} {\bf 75}: 762-777.   \\
Hayer A, Bhalla US (2005) Molecular switches at the synapse emerge from
receptor and kinase traffic. {\it PLoS Comput. Biol.} {\bf 1}: e20.  \\
Hill TL (1989) {\it Free Energy Transduction and Biochemical Cycle Kinetics}.
New York: Springer-Verlag.   \\
Honkura N, Matsuzaki M, Noguchi J, Ellis-Davies GCR, Kasai H (2008)
The subspine organization of actin fibers regulates the structure and
plasticity of dendritic spines. {\it Neuron} {\bf 57}: 719-729.  \\
Huganir RL, Nicoll RA (2013) AMPARs and synaptic plasticity: the last 25 years.
{\it Neuron} {\bf 80}: 704-717.   \\
Iossifov I, O'Roak BJ, Sanders SJ, Ronemus M, Krumm N, et al (2014)
The contribution of de novo coding mutations to autism spectrum disorder.
{\it Nature} {\bf 515}: 216-221.  \\
Jourdain P, Bergersen LH, Bhaukaurally K, et al (2007) Glutamate exocytosis
from astrocytes controls synaptic strength. {\it Nat. Neurosci.} {\bf 10}:
331-339.  \\
Kaizuka T, Takumi T (2018) Postsynaptic density proteins and their
involvement in neurodevelopmental disorders. {\it J. Biochem.}
{\bf 163}: 447-455.   \\
Kandel ER, Dudai Y, Mayford MR (2014) The molecular and systems biology
of memory. {\it Cell} {\bf 157}: 163-186.   \\
Karbowski J (2007) Global and regional brain metabolic scaling and its
functional consequences. {\it BMC Biol.} {\bf 5}: 18.   \\    
Karbowski J (2009) Thermodynamic constraints on neural dimensions,
firing rates, brain temperature and size.
{\it J. Comput. Neurosci.} {\bf 27}: 415-436.   \\
Karbowski J (2011) Scaling of brain metabolism and blood flow in relation
to capillary and neural scaling. {\it PLoS ONE} {\bf 6}: e26709.  \\
Karbowski J (2012) Approximate invariance of metabolic energy per synapse
during development in mammalian brains. {\it PLoS ONE} {\bf 7}: e33425.   \\
Karbowski J (2014) Constancy and trade-offs in the neuroanatomical and metabolic
design of the cerebral cortex. {\it Front. Neural Circuits} {\bf 8}: 9.  \\
Karbowski J (2015) Cortical composition hierarchy driven by spine proportion
economical maximization or wire volume minimization.  {\it PloS Comput. Biol. } 
{\bf 11}: e1004532.   \\
Kasai H, Matsuzaki M, Noguchi J, Yasumatsu N, Nakahara H (2003) 
Structure-stability-function relationships of dendritic spines.
{\it Trends Neurosci.} {\bf 26}: 360-368.   \\
Khakh BS, North RA (2012) Neuromodulation by extracellular ATP and P2X
receptors in the CNS. {\it Neuron} {\bf 76}: 51-69.  \\
Kim B, Hawes SL, Gillani F, Wallace LJ, Blackwell KT (2013) Signaling pathways
involved in striatal synaptic plasticity are sensitive to temporal pattern
and exhibit spatial specificity. {\it PloS Comput. Biol. }  {\bf 9}: e1002953.  \\
Kirkpatrick J, Pascanu R, Rabinowitz N, Veness J, Desjardins G, Rusu AA, et al
(2017) Overcoming catastrophic forgetting in neural networks.
{\it Proc. Natl. Acad. Sci. USA}  {\bf 114}: 3521-3526.   \\
Krebs EG (1981) Phosphorylation and dephosphorylation of glycogen phosphorylase:
a prototype for reversible covalent enzyme modification. {\it Curr. Top. Cell.
Regul.} {\bf 18}: 401-419.    \\
Laughlin SB, de Ruyter van Steveninck RR, Anderson JC (1998) The metabolic
cost of neural information. {\it Nature Neurosci.} {\bf 1}: 36-40.   \\
Lebowitz JL, Spohn H (1999) A Gallavotti-Cohen-type symmetry in the large
deviation functional for stochastic dynamics.
{\it J. Stat. Phys.}  {\bf 95}: 333-365.  \\
  Leibold C, Kempter R (2008) Sparseness constrains the prolongation of
  memory lifetime via synaptic metaplasticity. {\it Cereb. Cortex} {\bf 18}:
  67-77.   \\
Levy WB, Baxter RA (1996) Energy efficient neural codes. {\it Neural Comput.} 
{\bf 8}: 531-543.   \\
 Li J, Wilkinson B, Clementel VA, Hou J, O'Dell TJ, Coba MP (2016) Long-term
  potentiation modulates synaptic phosphorylation networks and reshapes the
  structure of the postsynaptic interactome. {\it Sci. Signal.} {\bf 9}: rs8.  \\
 Lin JW, Ju W, Foster K, Lee SH, Ahmadian G, Wyszynski M, Wang YT, Sheng M (2000)
 Distinct molecular mechanisms and divergent endocytotic pathways of AMPA receptor
 internalization. {\it Nat. Neurosci.} {\bf 3}: 1282-1290.  \\
Lisman J, Yasuda R, Raghavachari S (2012) Mechanisms of CaMKII action in long-term 
potentiation. {\it Nat. Rev. Neurosci.} {\bf 13}: 169-182.   \\
Logothetis NK (2008) What we can do and what we cannot do with fMRI.
{\it Nature} {\bf 453}: 869-878.  \\ 
 Matsuzaki M, Ellis-Davies GCR, Nemoto T, Miyashita Y, Iino M, Kasai H
(2001) Dendritic spine geometry is critical for AMPA receptor expression
in hippocampal CA1 pyramidal neurons. {\it Nat. Neurosci.} {\bf 4}: 1086-1092.  \\
Mehta P, Schwab DJ (2012) Energetic costs of cellular computation.
{\it Proc. Natl. Acad. Sci. USA} {\bf 109}: 17978-17982.  \\
Meyer D, Bonhoeffer T, Scheuss V (2014) Balance and stability
of synaptic structures during synaptic plasticity. {\it Neuron}
{\bf 82}: 430-443.  \\
Michalski PJ (2013) The delicate bistability of CaMKII. {\it Biophys. J.} {\bf 105}:
794-806.   \\
Miller P, Zhabotinsky AM, Lisman JE, Wang X-J (2005) The stability of a stochastic
CaMKII switch: Dependence on the number of enzyme molecules and protein turnover.
{\it PLoS Biol.} {\bf 3}: e107.   \\
Molden RC, Goya J, Khan Z, Garcia BA (2014) Stable isotope labeling of 
phosphoproteins for large-scale phosphorylation rate determination.
{\it Molecular \& Cellular Proteomics} {\bf 13}: 1106-1118.   \\
Montgomery JM, Madison DV (2004) Discrete synaptic states define a major
  mechanism of synapse plasticity. {\it Trends Neurosci.} {\bf 27}: 744-750.   \\
Nicolis G, Prigogine I (1977) {\it Self-Organization in Nonequilibrium Systems}.
Wiley: New York, NY.   \\
Nimchinsky EA, Yasuda R, Oertner TG, Svoboda K (2004) The number of glutamate
receptors opened by synaptic stimulation in single hippocampal spines.
{\it J. Neurosci.} {\bf 24}: 2054-2064.   \\
O'Connor DH, Wittenberg GM, Wang SSH (2005) Graded bidirectional synaptic plasticity
is composed of switch-like unitary events.
{\it Proc. Natl. Acad. Sci. USA} {\bf 102}: 9679-9684.   \\
Pankratov Y, Lalo U, Krishtal OA, Verkhratsky A (2009) P2X receptors and synaptic
plasticity. {\it Neuroscience} {\bf 158}: 137-148.   \\
Petersen CC, Malenka RC, Nicoll RA, Hopfield JJ (1998) All-or-none potentiation
at CA3-CA1 synapses. {\it Proc. Natl. Acad. Sci. USA} {\bf 95}: 4732-4737.  \\
Phillips R, Kondev J, Theriot J, Garcia H (2012) {\it Physical Biology of the Cell}.
Garland Science: London.    \\
Poo M-m, Pignatelli M, Ryan TJ, Tonegawa S, Bonhoeffer T, Martin KC, Rudenko A,
Tsai L-H, Tsien RW, Fishell G, et al (2016) What is memory? The present state of
the engram. {\it BMC Biol.} {\bf 14}: 40.   \\
Pougnet JT, Toulme E, Martinez A, Choquet D, Hosy E, Boue-Grabot E (2014)
ATP P2X receptors downregulate AMPA receptor trafficking and postsynaptic
efficacy in hippocampal neurons. {\it Neuron} {\bf 83}: 417-430.   \\
Qian H (2007) Phosphorylation energy hypothesis: Open chemical systems and
their biological function. {\it Annu. Rev. Phys. Chem.} {\bf 58}: 113-142.  \\
Qian H (2006) Open-system nonequilibrium steady state: Statistical thermodynamics,
fluctuations, and chemical oscillations. {\it J. Phys. Chem. B} {\bf 110}: 15063-15074.  \\
Raichle ME, Mintun MA (2006) Brain work and brain imaging. {\it Annu. Rev. Neurosci.}
{\bf 29}: 449-476.  \\
Rolfe DFS, Brown GC (1997) Cellular energy utilization and molecular origin of
standard metabolic rate in mammals. {\it Physiol. Revs.} {\bf 77}: 731-758.  \\
Sabatini BL, Oertner TG, Svoboda K (2002) The life cycle of Ca$^{2+}$ ions
in dendritic spines. {\it Neuron} {\bf 33}: 439-452.  \\
Schnakenberg J (1976) Network theory of microscopic and macroscopic behavior
of master equation systems. {\it Reviews of Modern Physics} {\bf 48}: 571-585.  \\
Sheng M, Hoogenraad CC (2007) The postsynaptic architecture of excitatory synapses:
A more quantitative view. {\it Annu. Rev. Biochem.} {\bf 76}: 823-847.  \\
Sheng M, Kim E (2011) The postsynaptic organization of synapses.
{\it Cold Spring Harb Perspect Biol} {\bf 3}: a005678.   \\
Shoenbaum G, Chiba AA, Gallagher M (1999) Neural encoding in orbitofrontal cortex
and basolateral amygdala during olfactory discrimination learning.
{\it J. Neurosci.} {\bf 19}: 1876-1884.   \\  
Shulman RG, Rothman DL (1998) Interpreting functional imaging studies in terms
of neurotransmitter cycling. {\it Proc. Natl. Acad. Sci. USA} {\bf 95}: 11993-11998.  \\
Shulman RG, Rothman DL, Hyder F (1999) Stimulated changes in localized cerebral energy
consumption under anesthesia. {\it Proc. Natl. Acad. Sci. USA} {\bf 96}: 3245-3250.  \\
Shulman RG, Rothman DL, Behar KL, Hyder F (2004) Energetic basis of brain activity:
implications for neuroimaging. {\it Trends Neurosci.} {\bf 27}: 489-495.  \\
Smith MA, Riby LM, van Eekelen JAM, Foster JK (2011) Glucose enhancement of human
memory: A comprehensive research review of the glucose memory facilitation effect.
{\it Neurosci. Biobehavior. Revs.} {\bf 35}: 770-783.  \\
Smolen P, Baxter DA, Byrne JH (2012) Molecular constraints on synaptic tagging
and maintenance of long-term potentiation: a predictive model.
{\it PLoS Comput. Biol.} {\bf 8}: e1002620.    \\
Sonnay S, Duarte JM, Just N, Gruetter R (2016) Compartmentalised energy metabolism
supporting glutamatergic neurotransmission in response to increased activity in
the rat cerebral cortex: A 13C MRS study in vivo at 14.1 T.
{\it J. Cereb. Blood Flow Metab.} {\bf 36}: 928.   \\
Sonnay S, Poirot J, Just N, Clerc AC, Gruetter R, Rainer G, Duarte JMN (2018)
Astrocytic and neuronal oxidative metabolism are coupled to the rate of
glutamate-glutamine cycle in the tree shrew visual cortex.
{\it Glia} {\bf 66}: 477-491.  \\
Star EN, Kwiatkowski DJ, Murthy VN (2002) Rapid turnover of actin in
dendritic spines and its regulation by activity. {\it Nature Neurosci.}
{\bf 5}: 239-246.   \\
Sun L, Zhang Y, Hwang G, Jiang J, Kim D, et al (2018) Synaptic computation
enabled by Joule heating of single-layered semiconductors for sound localization.
{\it Nano Lett.} {\bf 18}: 3229-3234.   \\
Takeuchi T, Duszkiewicz AJ, Morris RGM (2014) The synaptic plasticity and memory
hypothesis: encoding, storage and persistence.  {\it Phil. Trans. R. Soc. B}
{\bf 369}: 20130288.   \\
  Trinidad JC, Specht CG, Thaalhammer A, Schoepfer R, Burlingame AL (2006)
  Comprehensive identification of phosphorylation sites in postsynaptic density
  preparations. {\it Molecular \& Cellular Proteomics} {\bf 5}: 914-922.   \\
  Trinidad JC, Barkan DT, Gulledge BF, Thaalhammer A, Sali A, Schoepfer R,
  Burlingame AL (2012) Global identification and characterization of both
  O-GlcNAcylation and phosphorylation at murine synapse. 
{\it Molecular \& Cellular Proteomics} {\bf 11}: 215-229.   \\
  Van den Broeck C, Esposito M (2015) Ensemble and trajectory thermodynamics:
  A brief introduction. {\it Physica A} {\bf 418}: 6-16.   \\
  Visscher K, Schnitzer MJ, Block SM (1999) Single kinesin molecules studied
  with a molecular force clamp. {\it Nature}  {\bf 400}: 184-189.   \\
  Volgushev M, et al (2004) Probability of transmitter release at neocortical
  synapses at different temperatures. {\it J. Neurophysiol.}  {\bf 92}: 212-220.  \\
  Yoshimura Y, Yamauchi Y, Shinkawa T, Taoka M, et al (2004) Molecular constituents
  of the postsynaptic density fraction revealed by proteomic analysis using
  multidimensional liquid chromatography-tandem mass spectrometry. {\it J. Neurochem.}
  {\bf 88}: 759-768.    \\
  Zhang J, Petit CM, King DS, Lee AL (2011) Phosphorylation of a PDZ domain extension
  modulates binding affinity and interdomain interactions in postsynaptic density-95
  (PSD-95) protein, a membrane-associated guanylate kinase (MAGUK).
  {\it J. Biol. Chem.} {\bf 286}: 41776-41785.  \\
  Zhu J, Shang Y, Zhang M (2016) Mechanistic basis of MAGUK-organized complexes
  in synaptic development and signalling. {\it Nat. Rev. Neurosci.} {\bf 17}:
  209-223.   \\
  Zhu F, Cizeron M, Qiu Z, Benavides-Piccione R, Kopanitsa MV, Skene NG, Koniaris B,
  DeFelipe J, Fransen E, Komiyama NH, Grant SGN (2018) Architecture of the mouse
  brain synaptome. {\it Neuron} {\bf 99}: 781-799.   \\

\newpage

{\bf \large Figure Captions}

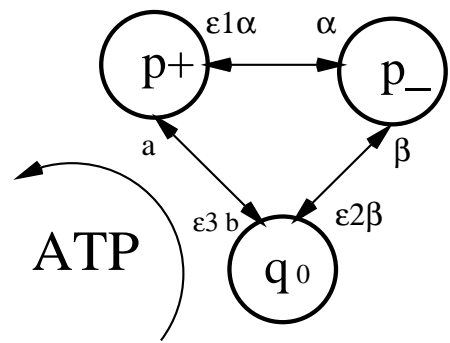
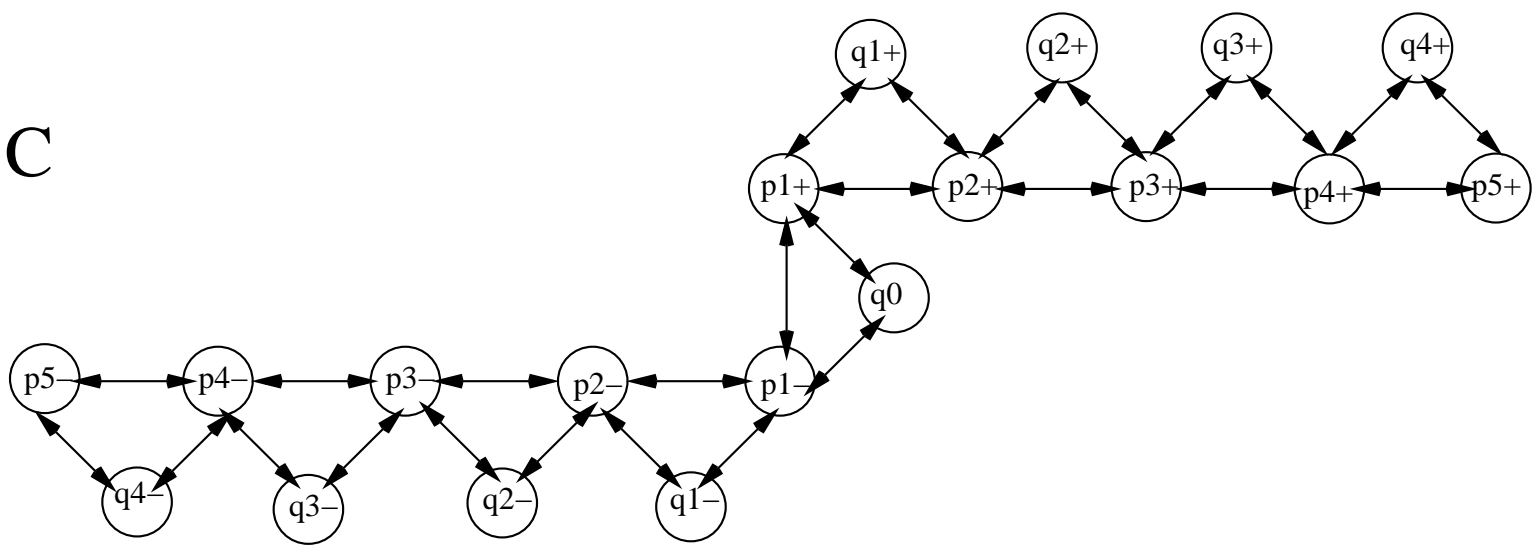
{\bf Fig. 1} \\
{\bf Cascade model of synaptic plasticity with cyclic and noncyclic reactions.}
(A) An example of the 3 state model with noncyclic reactions. For this model with
a ``ladder'' structure EPR at steady state is zero (see the Methods), and hence this
configuration is at thermal equilibrium with the environment and has a vanishing metabolic
rate. This is not a realistic situation.  \\
(B) Synaptic plasticity model with cyclic reactions. This model yields nonzero
EPR and metabolic rate at steady state due to ``reverberating'' loop that creates
a nonzero (cyclic) probability flux (see the Methods). State with the probability
$p_{-}$ describes a protein in a ground state, state with the probability $q_{0}$ corresponds
to the protein + substrate complex, and state with the probability $p_{+}$  is the activated
state of the protein + substrate complex that can signal (activate)
to other downstream molecules. To initiate some function a synapse must move from the
ground state $p_{-}$ to state $p_{+}$. However, direct transitions between $p_{-}$ and $p_{+}$
are rare (although possible due to thermal fluctuations), and the protein has to use an
alternative pathway to reach the activated state $p_{+}$. This involves a sequence of
two transitions from $p_{-}$ to $q_{0}$ (binding a protein with a substrate; relatively fast), 
and from $q_{0}$ to $p_{+}$ (activation of the protein + substrate complex). The latter
transition is powered by ATP hydrolysis (ATP $\leftrightarrow$ ADP + P), which
provides a necessary energy for speeding up this transition. When local concentration
of ATP increases due to calcium influx, the transition rate $q_{0} \rightarrow p_{+}$
increases accordingly. The process $q_{0} \rightarrow p_{+}$ can be thought as protein
phosphorylation, e.g., phosphorylation of CaMKII and/or PSD-95, which are the most
abundant and one of the main signaling molecules in dendritic spines (Sheng and Kim 2011).
This cyclic molecular motif is known as a phosphorylation-dephosphorylation cycle and
serves as a building block for constructing more complex signaling molecular networks
(Hill 1989; Qian 2007).     \\
(C) Cascade synaptic model with many basic cyclic motifs. This is an expansion
of the model in panel B, and contains two chains of ``up'' and ``down'' 
synaptic states corresponding to binary synaptic weight or a number of AMPA
receptors expressed in the spine membrane (Petersen et al 1998; O'Connor et al 2005).
Occupancy probabilities in the up and down states are respectively $p_{k}^{+}, q_{k}^{+}$ 
and $p_{k}^{-}, q_{k}^{-}$. The transitions $p_{k}^{\pm} \leftrightarrow p_{k+1}^{\pm}$
are spontaneous and correspond to the direct rare transitions $p_{-} \leftrightarrow p_{+}$
in panel B. The pathway 
$p_{k}^{\pm} \leftrightarrow q_{k}^{\pm} \leftrightarrow p_{k+1}^{\pm}$
is indirect, since it involves an intermediate state $q_{k}^{\pm}$ (corresponding to
state $q_{0}$ in panel B). The reactions $q_{k}^{\pm} \rightarrow p_{k+1}^{\pm}$
correspond to protein phosphorylation driven by ATP hydrolysis with transition
rates $a_{k}^{\pm}$, while the reverse reactions $p_{k+1}^{\pm} \rightarrow q_{k}^{\pm}$
describe much slower dephosphorylations with transition rate $\epsilon_{3}b$ that is
a few orders of magnitude smaller than $a_{k}^{\pm}$ (Qian 2006, 2007). All the
transition rates $a_{k}^{\pm}$ are proportional to the amplitude $a_{0}$.
The transitions in the direction $p_{k}^{+} \rightarrow p_{k+1}^{+}$ (plus
$p_{1}^{-} \rightarrow p_{1}^{+}$) of the up states are associated with LTP process,
whereas the transitions in the direction $p_{k}^{-} \rightarrow p_{k+1}^{-}$ of the down
states are related to LTD. The dynamics of downstream cycles (higher index $k$)
both for the up and down states are scaled down by a factor $e^{-kz}$ with a characteristic
slowing down factor $z$, which means that the transitions within deeper states are
progressively slower. When the synapse is stimulated and
a plasticity process initiated, only the ATP-driven rates from $q_{k}^{\pm}$ to
$p_{k+1}^{\pm}$, denoted as $a_{k}^{\pm}$ are time dependent. Specifically, for 
synapses undergoing LTP the rates $a_{k}^{+}$ are time dependent, while for
synapses undergoing LTD the rates $a_{k}^{-}$ change in time. The state with
the occupancy $p_{1}^{-}$ is the ground state (depressed).

\vspace{0.3cm}

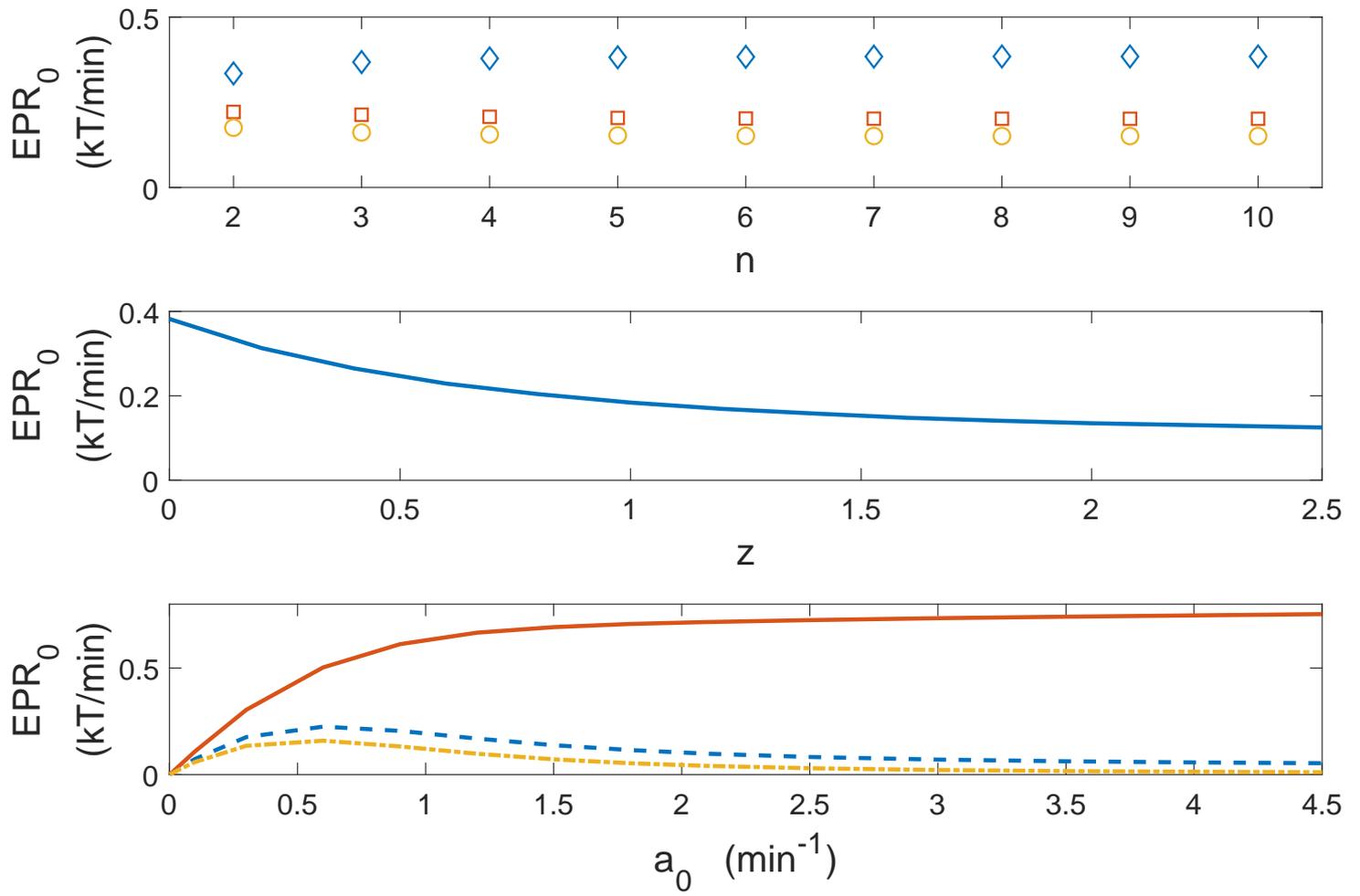
{\bf Fig. 2} \\
{\bf Dependence of baseline synaptic plasticity energy rate EPR$_{0}$ on
  internal synaptic parameters.}
Baseline EPR$_{0}$ is almost independent on the number of basic synaptic states
$n$ (upper panel; blue diamonds for $z=0$, red squares for $z=0.8$, yellow
circles for $z=1.5$), but it decreases monotonically with increasing the
slowing-down rate $z$ (middle panel).
Dependence of EPR$_{0}$ on the ATP-driven transition rate
$a_{0}$ is more complex (bottom panel): when $z=0$ (solid line) EPR$_{0}$ increases
monotonically with $a_{0}$ up to a saturation, while for $z > 0$ (dashed line
for $z=0.8$, dashed-dotted line for $z=1.5$) EPR$_{0}$ exhibits a bimodal
shape and decays to 0 for large $a_{0}$.

\vspace{0.3cm}

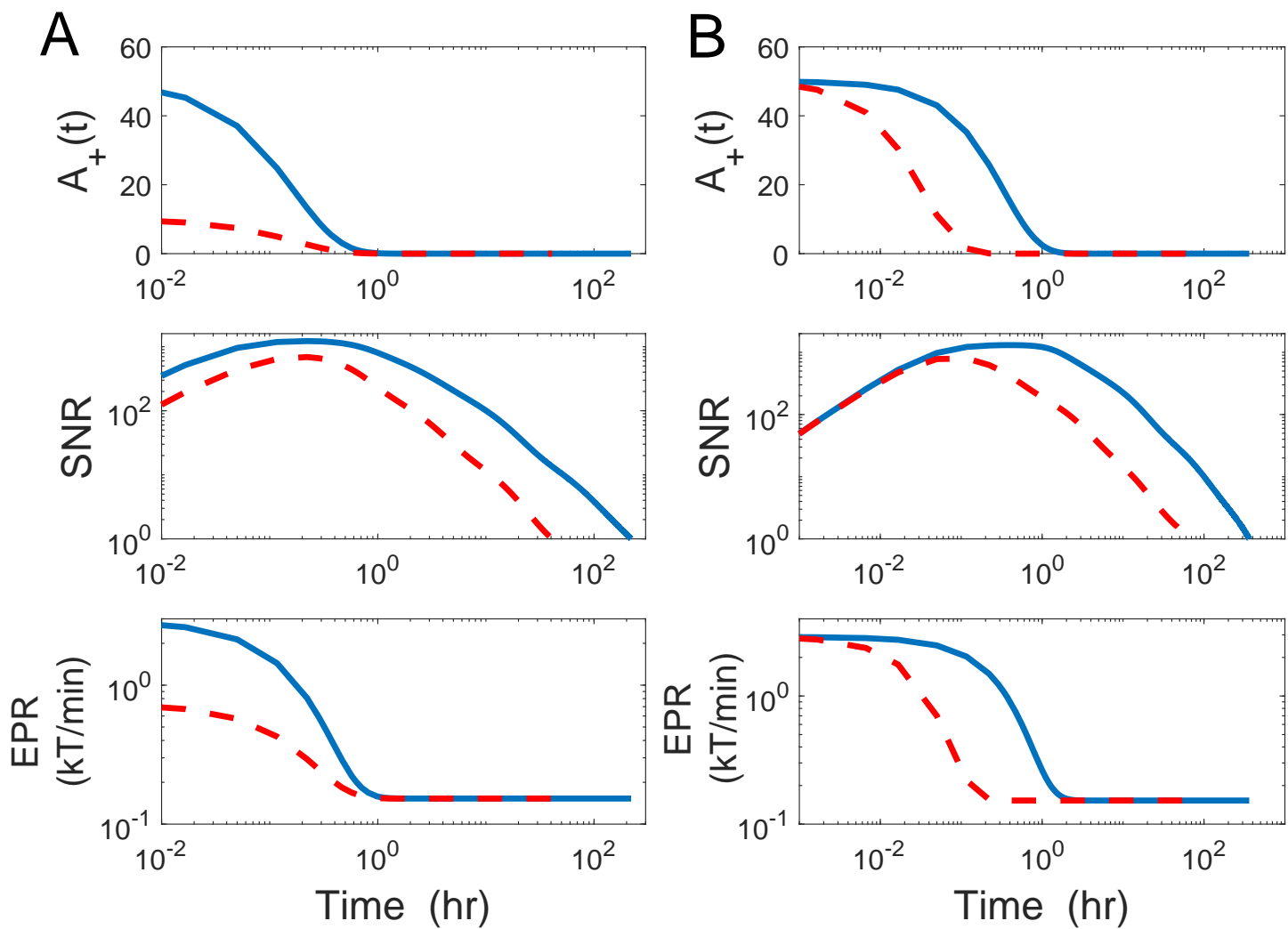
{\bf Fig. 3} \\
{\bf Time course of memory trace SNR and associated entropy production rate
  EPR after synaptic stimulation}.
Synaptic stimulation $A_{\pm}(t)$ affects the ATP driven transitions $a^{\pm}_{k}$ in the
following way: $a^{\pm}_{k} \mapsto a^{\pm}_{k}(t)= a^{\pm}_{k}[1 + A_{\pm}(t)]$, where
$A_{\pm}(t)= A_{\pm}\exp(-t/\tau)$.
A) $A_{+}(t)$, SNR, and EPR as functions of time for different stimulus amplitudes
$A_{+}$ (solid blue line for $A_{+}= 50$, dashed red line for $A_{+}= 10$).
B) $A_{+}(t)$, SNR, and EPR as functions of time for different stimulus durations
$\tau$ (solid blue line for $\tau= 20$ min, dashed red line for $\tau= 2$ min).
Note that for both cases A) and B) the SNR decays to zero as approximately a power law
$\sim t^{-\delta}$ with $\delta\approx 1.3$. This is much slower decay than EPR relaxation
to its baseline, which essentially follows the stimulation $A_{+}(t)$.
Moreover, EPR stabilizes at a nonzero value, which is a signature of a non-equilibrium
steady state with a corresponding nonzero metabolic rate. For all plots $z=1.5$.

\vspace{0.3cm}

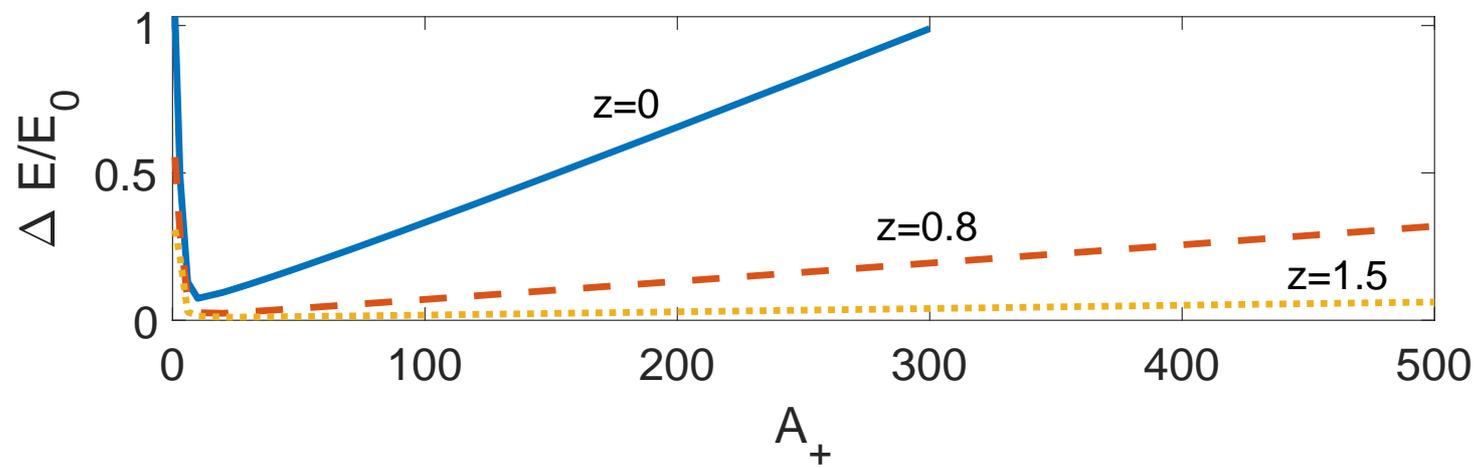
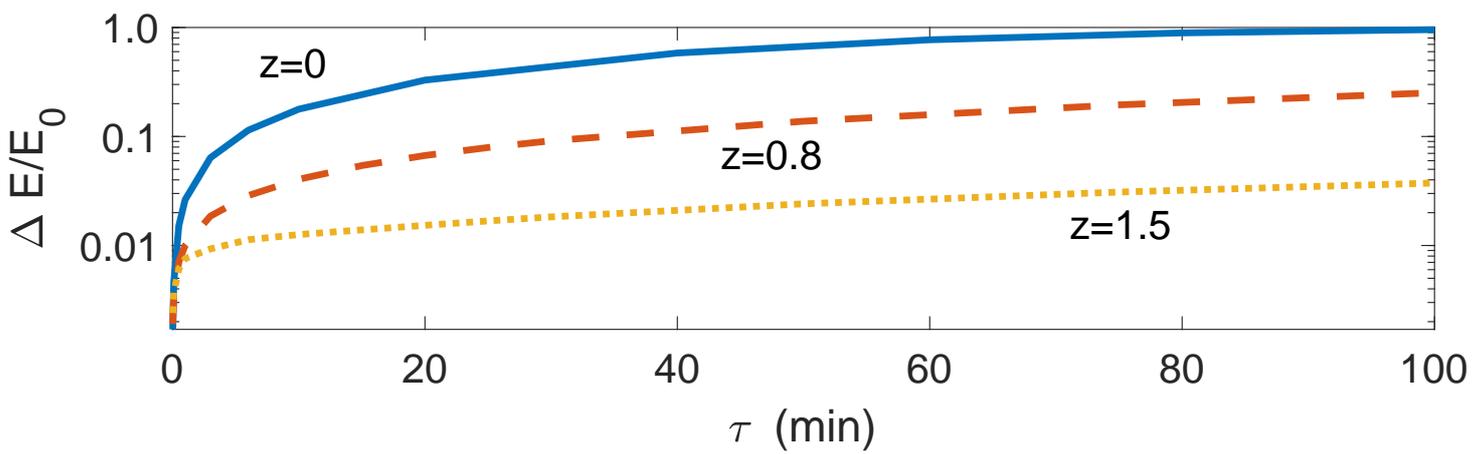
{\bf Fig. 4} \\
{\bf Relative cost of maintaining a memory trace can be small for progressively
  slower downstream synaptic transitions.}
Relative energy associated with a memory trace above baseline $\Delta E/E_{0}$
(where $\Delta E= E - E_{0}$, and $E_{0}=\mbox{EPR}_{0}T_{m}$) as a function of amplitude
$A_{+}$ and duration $\tau$ of the stimulation. Note that $\Delta E/E_{0}$ generally
gets smaller for increasing the rate of synaptic slowing down $z$.

\vspace{0.3cm}

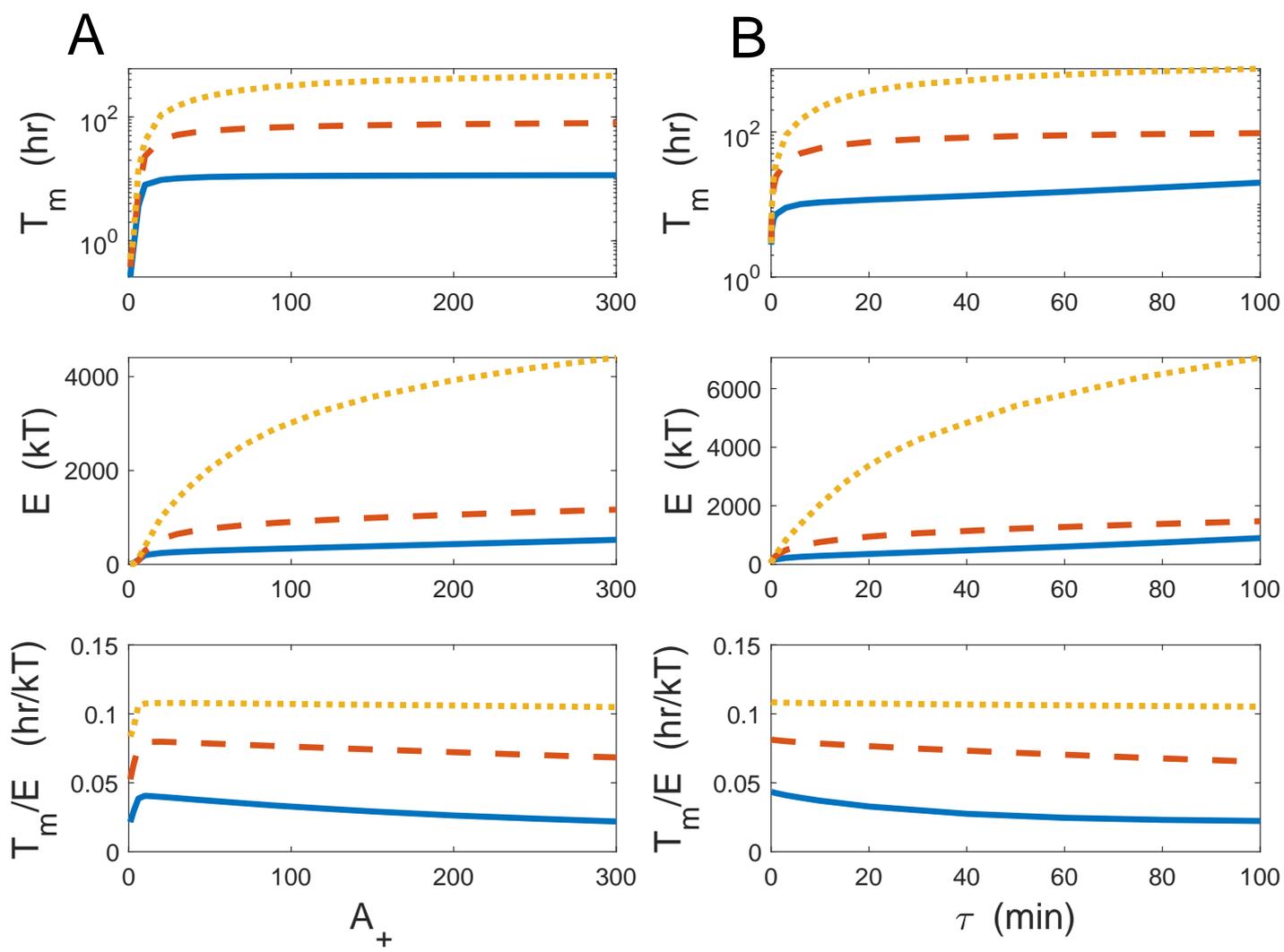
{\bf Fig. 5} \\
{\bf Effect of synaptic stimulation magnitude on memory lifetime and its energy cost.}
(A) Memory lifetime $T_{m}$, its energy cost $E$ and their ratio $T_{m}/E$ as functions
of stimulation amplitude $A_{+}$. (B) The same variables as functions of stimulus duration
$\tau$. Note that in both cases the ratio $T_{m}/E$ is essentially constant, which indicates
that longer memories need proportionally more energy. In all panels, solid blue line correspond
to $z=0$, dashed red line to $z=0.8$, and dotted yellow line to $z=1.5$.

\vspace{0.3cm}

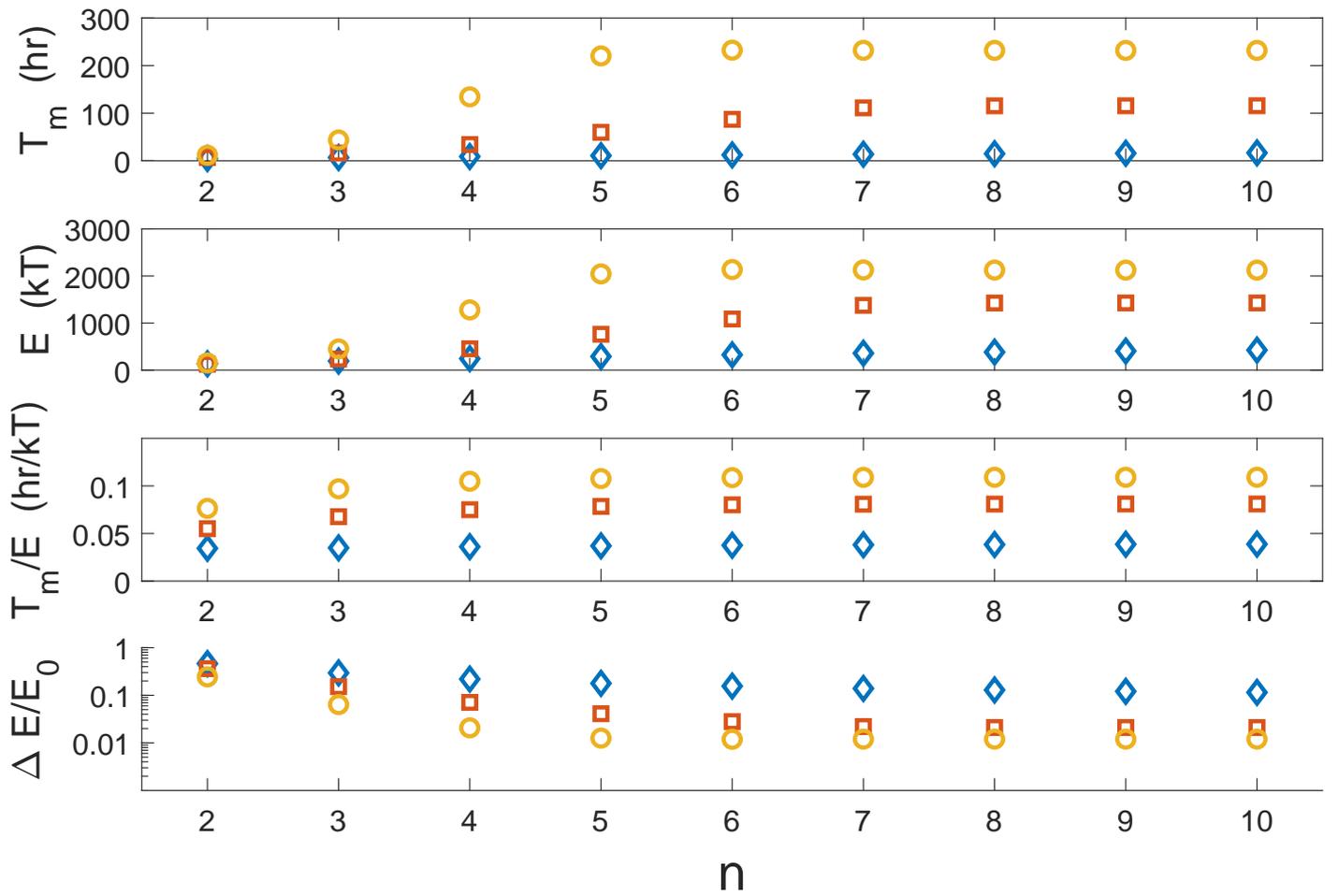
{\bf Fig. 6} \\
{\bf Memory lifetime and its energy cost as functions of the number of
  synaptic states $n$}. Memory lifetime per energy is essentially constant
as a function of $n$. Note that the relative energy above baseline for maintaining
memory trace is a tiny percentage for $z > 0$ and larger $n$.
Blue diamonds correspond to $z=0$, red squares to $z=0.8$, and yellow circles
to $z=1.5$.

\vspace{0.3cm}

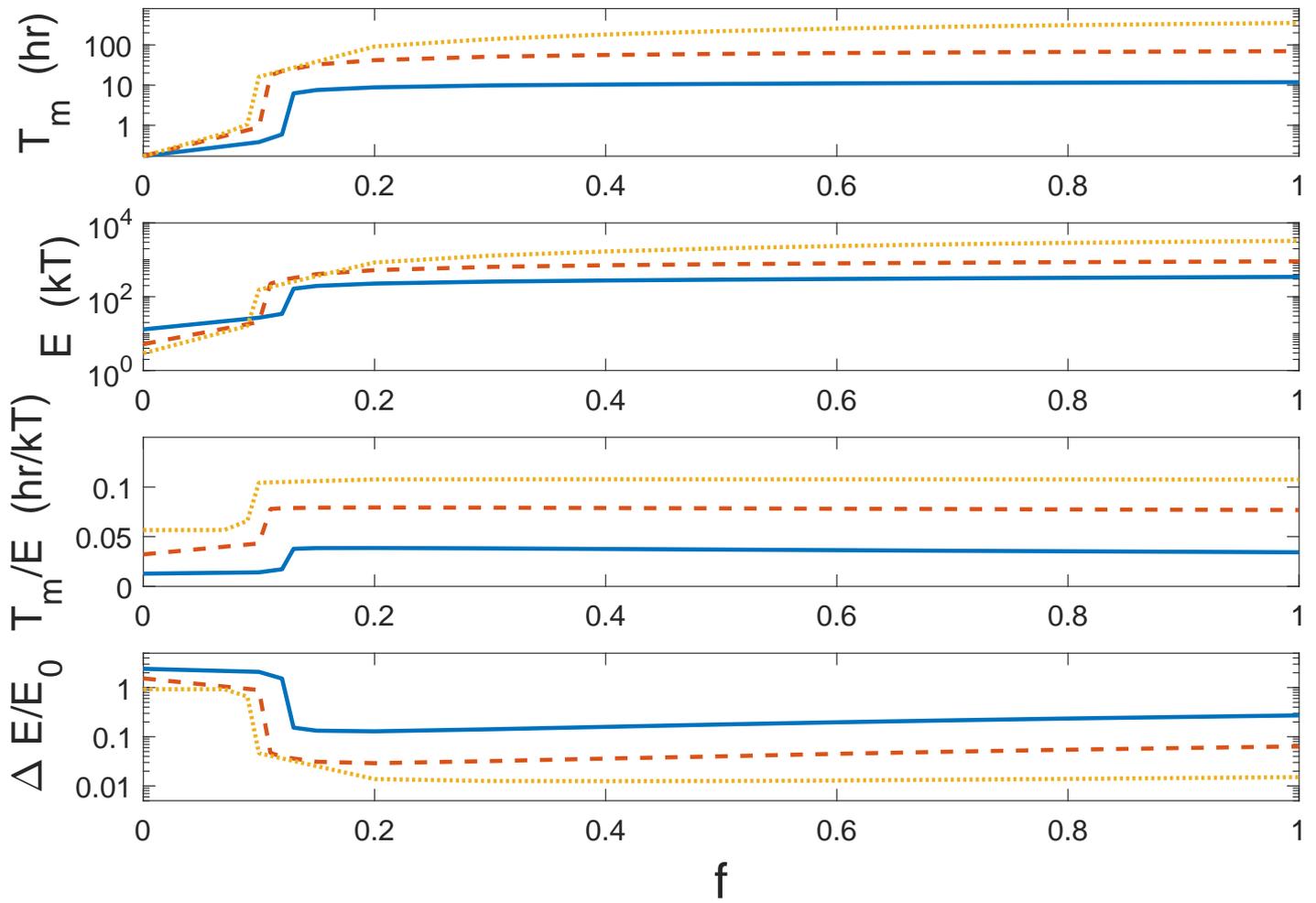
{\bf Fig. 7} \\
{\bf Dependence of memory lifetime and its energy cost on the fraction
  of potentiated synapses $f$}. For sufficiently large $f$ ($f > 0.2$), memory lifetime
$T_{m}$, its energy cost $E$, as well as $T_{m}/E$ and $\Delta E/E_{0}$ are all essentially
constant. Blue solid line corresponds to $z=0$, red dashed line to $z=0.8$,
and yellow dotted line to $z=1.5$.

\vspace{0.3cm}

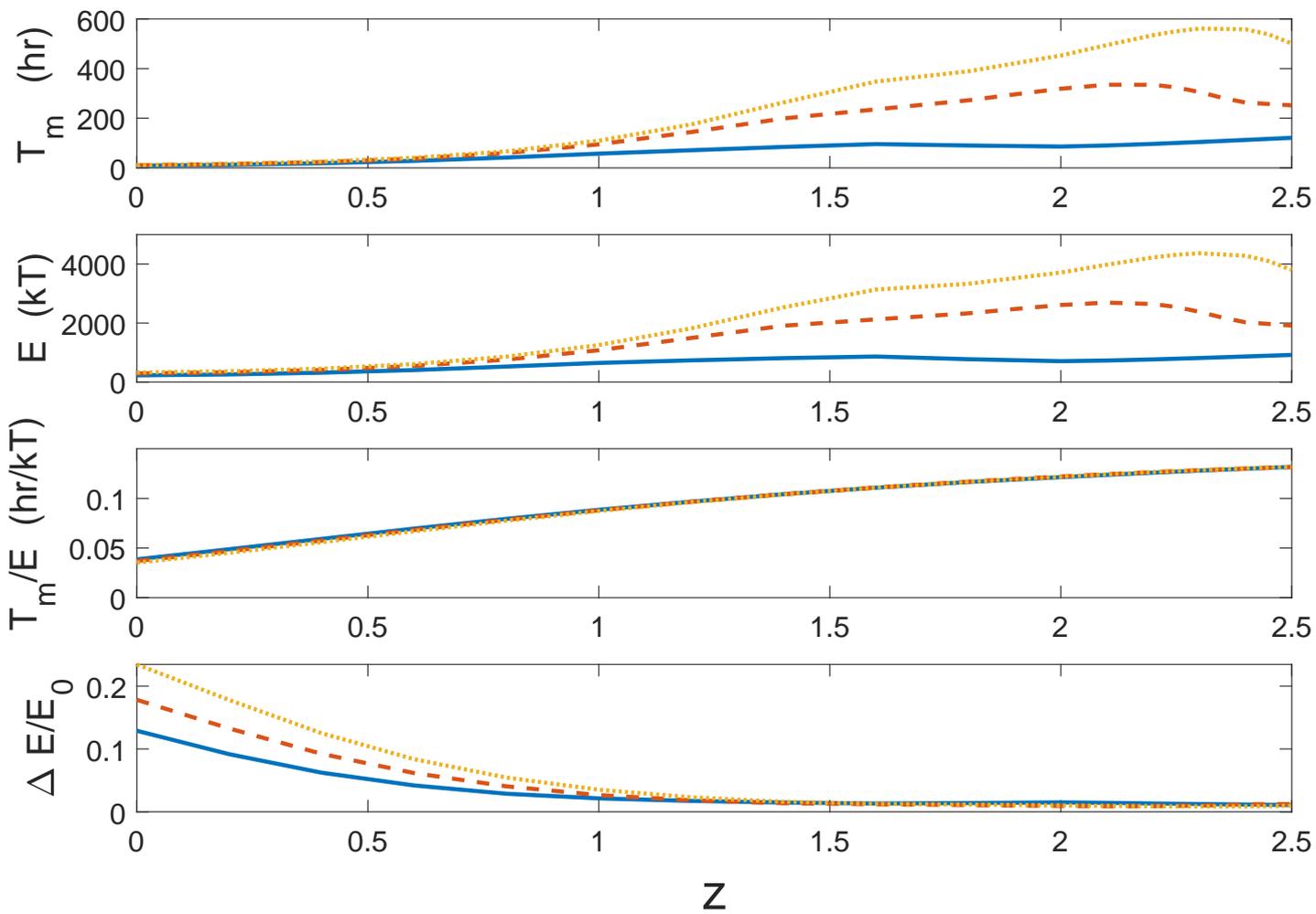
{\bf Fig. 8} \\
{\bf Memory lifetime and its energy cost as functions of slowing-down
  rate $z$}. The ratio $T_{m}/E$ increases monotonically but weakly with $z$,
and $\Delta E/E_{0}$ decreases with $z$.
Blue solid line corresponds to $f=0.2$, red dashed line to $f=0.5$, and
yellow dotted line to $f=0.8$.

\vspace{0.3cm}

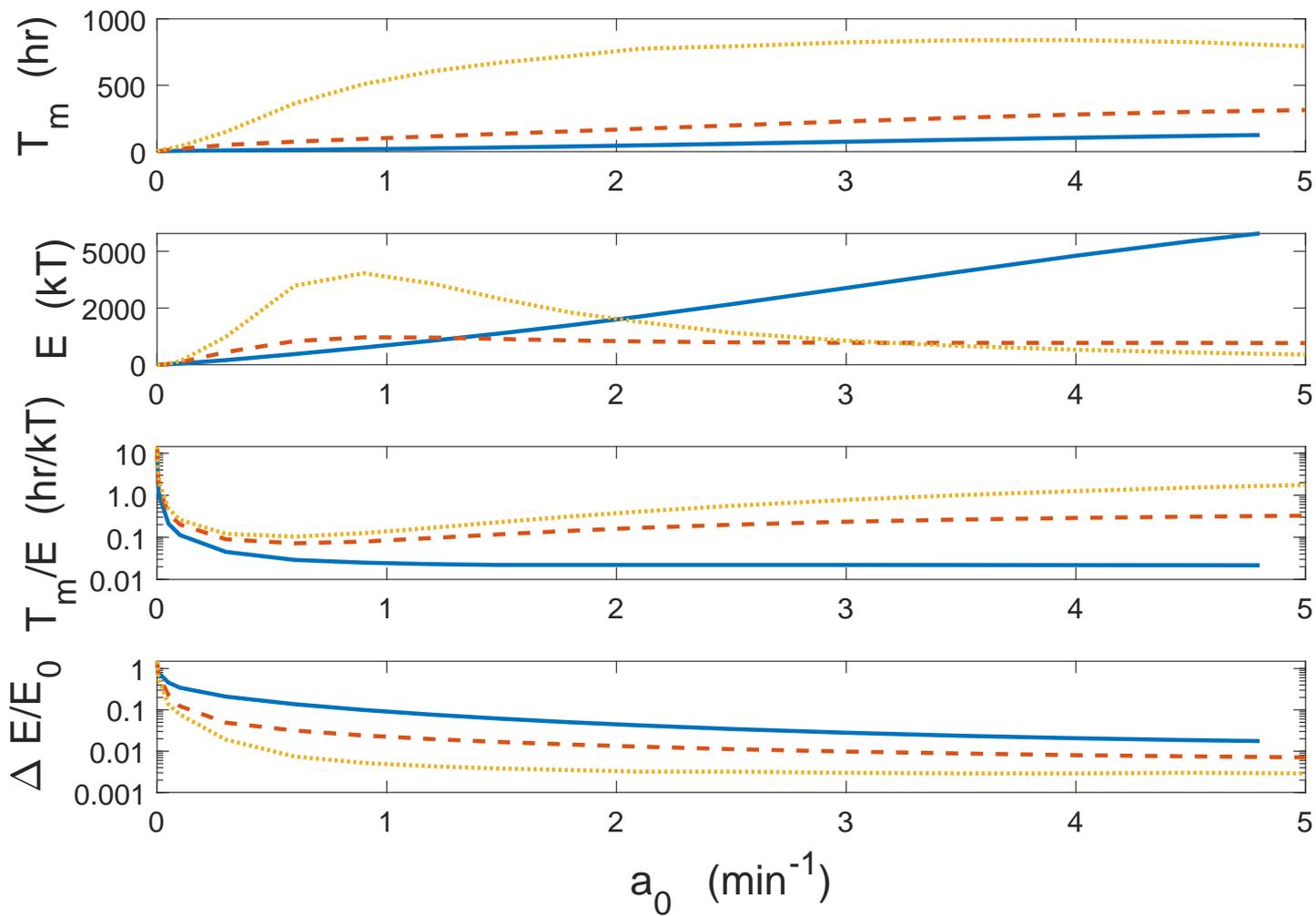
{\bf Fig. 9} \\
{\bf Memory lifetime and its energy cost as functions of the phosphorylation
  rate amplitude $a_{0}$}. Surprisingly, for sufficiently large $z$ longer memories can
require less energy (two upper panels).
The ratio $T_{m}/E$ exhibit a minimum for some $a_{0}$, but increases relatively fast
for larger $a_{0}$, especially for larger $z$.
Blue solid line corresponds to $z=0$, red dashed line to $z=0.8$,
and yellow dotted line to $z=1.5$.

\newpage

\begin{table}
\begin{center}
  \caption{ Energetics of the major molecular processes involved in dendritic spine
    plasticity for rat brain.} 

\begin{tabular}{|l l|}
\hline

Molecular             &        ATP rate                  \\
mechanism             &        (min$^{-1}$)                \\

\hline

{\bf Extra-synaptic}      &                                      \\
                                  &                                    \\
$\;$ Glutamate recycling       &            1602                          \\
                                    &                                    \\
$\;$ ATP binding to spine       &           60                         \\
                                &                                    \\
{\bf Intra-synaptic}      &                                      \\
                               &                                    \\
$\;$ Ca$^{2+}$ removal       &                6000                    \\
                              &                                    \\
$\;$ Protein phosphorylation:      &                                 \\
 $\;$ $\;$ active LTP$^{*}$       &            $310000-910000$         \\
$\;$ $\;$ (resting non-LTP)       &           (7500)                 \\

                          &                                    \\
$\;$ Protein synthesis     &              3700               \\
                             &                                    \\
$\;$ Actin treadmilling        &             8000             \\
                           &                                     \\
$\;$ Receptor trafficking:        &                                      \\
 $\;$ $\;$ movement    &              8184                  \\  
 $\;$ $ \;$ insertion   &             18.7                \\
                           &                                    \\
{\bf Modulatory}      &                                      \\
                          &                                    \\
$\;$ P2X receptors       &            2871                   \\

\hline

\hline
\end{tabular}

$^{*}$ Values depend on Ca$^{2+}$ stimulation rate.

\end{center}

\end{table}

\newpage

\begin{table}
\begin{center}
  \caption{ Comparison of the metabolic costs of synaptic plasticity and fast excitatory
    synaptic transmission for rat brain.} 

\begin{tabular}{|l l l|}
\hline

Synaptic plasticity         &   Postsynaptic current   &  Plasticity/Transmission   \\
total cost (ATP/min)        &    cost (ATP/min)        &     relative cost ($\%$)          \\

\hline

                             &                         &                          \\
 $(3.4-9.4)\cdot 10^{5}$      &   $8.4\cdot 10^{6}$     &    $4.0-11.2$            \\
 (enhanced phosphorylation)  &                         &                           \\

\hline

\hline
\end{tabular}

\end{center}

The cost of the postsynaptic current (synaptic transmission) is estimated based on
Attwell and Laughlin (2001).

\end{table}

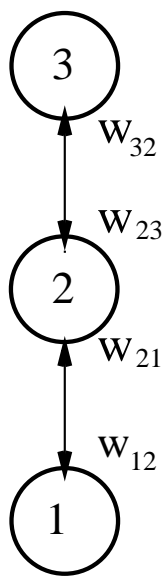
\end{document}